\documentclass[a4paper,11pt]{article}
\pdfoutput=1
\usepackage{jheppub}
\usepackage{amsmath}
\usepackage{slashed}
\usepackage{amssymb}
\usepackage{color}
\usepackage{graphicx}
\usepackage{hyperref}
\usepackage{xspace,slashed}
\usepackage[utf8]{inputenc}
\usepackage{subcaption}
\usepackage{multirow}
\usepackage{algorithm}
\usepackage{algorithmic}
\usepackage{stackengine}
\usepackage{enumitem}
\usepackage{etoolbox}

\usepackage{tikz}
\usepackage{tikz-feynman}

\newtoggle{draft}

\togglefalse{draft}

\iftoggle{draft} {
  \usepackage{changes}
}{%
\usepackage[final]{changes}
}

\definechangesauthor[name=pn, color=blue]{pn}
\definechangesauthor[name=km, color=magenta]{km}

\usepackage{booktabs}

\makeatletter
\g@addto@macro\bfseries{\boldmath}
\makeatother

\newcommand{\defargs}{} 

\newcommand{\barhatN}{{\bar{\hatN}}}

\newcommand{\hatone}{{\bf 1}}
\newcommand{\hatN}{{\bf N}}
\newcommand{\hatS}{{\bf S}}
\newcommand{\ptop}{p_t}
\newcommand{\hatptop}{{\slashed p}_t}
\newcommand{\hatpb}{{\slashed p}_b}
\newcommand{\tildeptop}{{\tilde p}_t}
\newcommand{\dtop}{d_t}
\newcommand{\mt}{m_t}
\newcommand{\tildemt}{{\tilde m}_t}
\newcommand{\qtop}{q_t}
\newcommand{\hatqtop}{{\slashed q}_t}
\newcommand{\hatk}{{\slashed k}}

\newcommand{\pu}{p_u}
\newcommand{\pd}{p_d}
\newcommand{\pb}{p_b}
\newcommand{\tildepd}{{\tilde p}_d}
\newcommand{\qd}{q_d}
\newcommand{\db}{d_b}
\newcommand{\as}{\alpha_s}
\newcommand{\gs}{g_s}
\newcommand\nf{n_{f}}
\newcommand\Tf{T_{F}}

\newcommand\Cf{C_F}

\newcommand{\mathd}{\mathrm{d}}

\newcommand{\msbar}{$\overline{\text{MS}}$}

\newcommand{\be}{\begin{equation}}
\newcommand{\ee}{\end{equation}}
\newcommand{\ep}{\epsilon}

\usepackage{soul}
\allowdisplaybreaks

\definecolor{azure}{rgb}{0.0, 0.9, 1.0}

\newcommand{\TTPaff}{Institute for Theoretical Particle Physics,
  KIT, 76128 Karlsruhe, Germany}
\newcommand{\MILaff}{INFN, Sezione di Milano-Bicocca, and Universit\`a di Milano-Bicocca, Piazza della Scienza 3, 20126 Milano, Italy}
\newcommand{\MPIaff}{ Max-Planck-Institut für Physik, 80805 M\"unchen, Germany}
\newcommand{\Saclay}{Universit\'e Paris-Saclay, CNRS, IJCLab, 91405 Orsay, France}

\preprint{
  \begin {flushright}
    TTP23-003, P3H-23-007 
  \end{flushright}
}

\title{
Linear power corrections to single top production processes at the LHC
}
\author[a]{Sergei Makarov,}
\author[a]{Kirill Melnikov,}
\author[b,d]{Paolo Nason,}
\author[c,a]{Melih A. Ozcelik}
\affiliation[a]{\TTPaff}
\affiliation[b]{\MILaff}
\affiliation[c]{\Saclay}
\affiliation[d]{\MPIaff}

\emailAdd{sergei.makarov@kit.edu}
\emailAdd{kirill.melnikov@kit.edu}
\emailAdd{paolo.nason@mib.infn.it}
\emailAdd{melih.ozcelik@ijclab.in2p3.fr}

\abstract{ We discuss the linear power corrections to the electroweak
  production of top quarks at the LHC using renormalon calculus.  We
  show how such non-perturbative corrections can be obtained using the
  Low-Burnett-Kroll theorem, which provides the first subleading term
  to the expansion of the real-emission amplitudes around the soft
  limit. We demonstrate  that there are no linear power corrections to the
  total cross sections of arbitrary processes of a 
 single top production type  provided
  that these cross sections are expressed in terms of a short-distance top quark 
  mass.  We also derive a universal formula for the linear power
  corrections to generic observables that involve the top-quark
  momentum.  }

\begin{document}

\maketitle 
\section{Introduction}

High rates and clean signatures of top quark production processes at
the LHC have ushered the era of high-precision exploration of top
quark properties. Such studies can be performed in processes where top
quarks and anti-top quarks are produced in pairs via strong
interactions, and also in processes where \emph{single} top quarks are
produced  by flavor-changing electroweak charged currents.

Among many interesting quantities that one can study in such
processes, the top quark mass plays a particularly important role.
Experimentally, the top quark mass is already measured with very high
precision  and further improvements are expected at the
high-luminosity LHC.\footnote{For a recent review of top quark physics, including the mass measurements and the discussion of future 
  prospects, see ref.~\cite{Schwienhorst:2022yqu}.}
  Theoretically, there is a debate about
non-perturbative effects that affect \emph{all} existing top quark
measurements and require better understanding.

In the past, many of these discussions were framed as a dispute about
the type of mass that is best extracted from a particular
measurement.\footnote{An account of the different point of views with the
associated references is given in section 6.5.1 of
ref.~\cite{Azzi:2019yne}.}
It was sometimes argued that short-distance mass renormalisation schemes, for example  the
\msbar{} scheme, are preferable over the pole-mass scheme because
the pole mass is affected by infrared renormalons~\cite{Bigi:1994em,Beneke:1994sw}.
Studies of the apparent convergence of the perturbative expansion
in different mass schemes have also been performed to support these
arguments~\cite{Langenfeld:2009wd,Dowling:2013baa,Makela:2023wbk}.

However, it  is far from obvious  that the top quark mass renormalon is the 
only renormalon that affects  top quark production, in spite of being the
one that has attracted most attention.
Moreover, since no first-principles understanding of the  non-perturbative
effects in hadron collider processes currently exists, ultra-precise
determinations of many fundamental parameters at the LHC, including the
top quark mass, remain  obscure.

A possible step towards a better understanding of the non-perturbative
contributions to relevant LHC processes, including heavy quark
production, is to study power corrections using renormalon
calculus.\footnote{For recent applications see
  refs.~\cite{Caola:2021kzt,FerrarioRavasio:2018ubr}.} %
This technique works under the assumption that the renormalon
contributions are dominated by the large value of $b_0$, the
coefficient of the leading term of the QCD $\beta$-function.  More
specifically, one starts  with   a model theory with a large 
negative value of massless quark species $\nf$, and considers only the dominant terms in the
perturbative expansion, proportional to powers of $\as \nf$. In this
limit the coefficient of the leading term of the beta function equals
$b_{0,\nf}=-4 \Tf \nf/(12\pi)$;   it is  positive for negative
$\nf$, so that the model theory is asymptotically free. At the end of
the calculation one replaces $b_{0,\nf}$ with $b_0$-value in 
QCD, $b_0 = ( 11 C_A - 4 T_F n_f)/(12 \pi)$. 

It turns out that the results in the large-$b_0$ approximation  can be easily obtained 
from calculations in QCD  where the gluon carries a small mass $\lambda$. It can be shown that,
if an observable is linearly sensitive to $\lambda$,  there is a
renormalon in the perturbative expansion of this observable associated with a power
correction of order $\Lambda_{\rm QCD}$.  This procedure is well known, and it
has been reviewed in ref.~\cite{Beneke:1998ui}, where many
applications are also discussed.  A complete account of  how these
calculations are carried out, also including the contribution of
non-inclusive real corrections, is given in  Appendix~B of
ref.~\cite{FerrarioRavasio:2018ubr}.

Unfortunately, the application of the renormalon calculus is currently
limited to processes where no gluons appear in the   Feynman diagrams
that contribute at the leading order.  This feature prevents us from
applying the renormalon analysis to studying  non-perturbative effects in
the top quark \emph{pair production} process.  However, the
non-perturbative contributions to the $t$-channel single top production
process can be analysed using the renormalon calculus, since at the leading
order this process is a flavor-changing quark-quark scattering
mediated by the exchange of a $W$-boson.

We will show that such non-perturbative 
contributions can be determined for a class of processes
$pp \to t + X + q$, where $X$ is an arbitrary collection of colourless
particles, using the so-called Low-Burnett-Kroll (LBK) theorem
\cite{Low:1958sn,Burnett:1967km},\footnote{%
  For recent literature on the LBK theorem see
  ref.~\cite{Engel:2021ccn} and references therein.}  which  allows one
to obtain the \emph{first sub-leading contribution} to    the  expansion of 
the scattering amplitude for soft radiation.  Following the logic  of the LBK theorem, we will also be able to compute the
renormalon structure of the virtual corrections to the same generic
process.\footnote{We note that the connection between linear power corrections, soft radiation
and the LBK theorem was pointed out  a long time ago in refs.~\cite{Akhoury:1996ks,Akhoury:1997pb}.}

The rest of the paper is organised as follows. In the next section we
discuss the real emission contribution to the process
$pp \to t + q + g + X$ and explain how the ${\cal O}(\lambda)$ corrections
to the fully-differential partonic cross section can be computed using
the Low-Burnett-Kroll theorem. In Section~\ref{sect:virt} we
generalise this result to the computation of the virtual
corrections. In Section~\ref{sect:renorm} we combine the virtual
corrections with various renormalisation contributions. In
Section~\ref{sect:mass} we explain how to compute the change in the
cross section due to a top quark mass redefinition. In
Section~\ref{sect:final} we combine the various contributions and show
that the linear ${\cal O}(\Lambda_{\rm QCD})$ power corrections cancel in
the total cross section provided that a short-distance top-quark mass scheme 
is used. In Section~\ref{sect:altself} we illustrate an alternative way
to compute the effect of the self-energy insertions in  the external top
line, that allows one to perform the calculation directly in any
short-distance mass scheme.
In Section~\ref{sect:kinematics} we describe the computation of the ${\cal O}(\Lambda_{\rm QCD})$
corrections to observables that depend on the top quark momentum; at  variance with the total cross section,
we find that there \emph{are} linear power corrections to such observables. 
We present our conclusions  in 
Section~\ref{sect:concl}.
In Appendix~\ref{app:integrals} we provide  results for real and virtual integrals that we have used in the calculation,
while in Appendix~\ref{app:SL} we show how to reproduce the well-known result~\cite{Bigi:1994em,Beneke:1994bc} on the absence of
the ${\cal O}(\Lambda_{\rm QCD})$ corrections to
semileptonic decays of a heavy quark  using our technique.

\section{Real emission contribution to single top production  and the Low-Burnett-Kroll theorem}
\label{sect:real}
We consider the process of $t$-channel single top production in
association with a colourless system $X$
\begin{equation}
  u(\pu) + b(\pb) \to d (\pd) + t(\ptop) + X(p_X),
  \label{eq1.1a}
\end{equation}
and write the kinematics for the real correction to this process due to the emission of a massive gluon as follows
\begin{equation}
 u(\pu) + b(\pb) \to d (\qd) + t(\qtop) + X(p_X) + g(k).
\label{eq8.1}
\end{equation}
We note that we have used  different notations for the four-momenta  of the top quark and the down quark in the two cases.
This is done for   future convenience  since, as we will see, 
these momenta will absorb the recoil  due to  the  emitted    soft gluon.

\begin{figure}
    \centering
    \begin{subfigure}[t]{0.49\textwidth}
        \centering
        \vskip 0pt
        \includegraphics{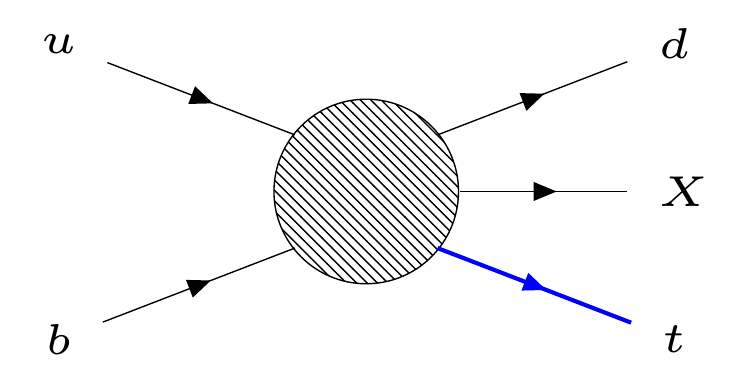}
    \end{subfigure}
    \begin{subfigure}[t]{0.49\textwidth}
        \centering
        \vskip 0pt
        \includegraphics{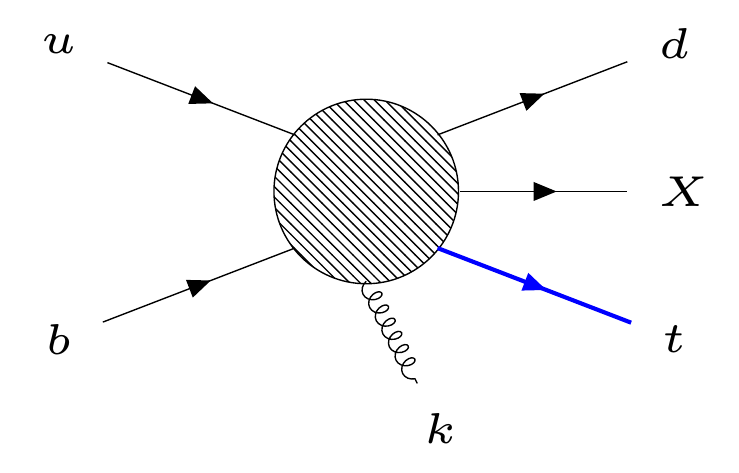}
    \end{subfigure}

    \begin{subfigure}[t]{0.49\textwidth}
        \centering
        \vskip 0pt
        \includegraphics{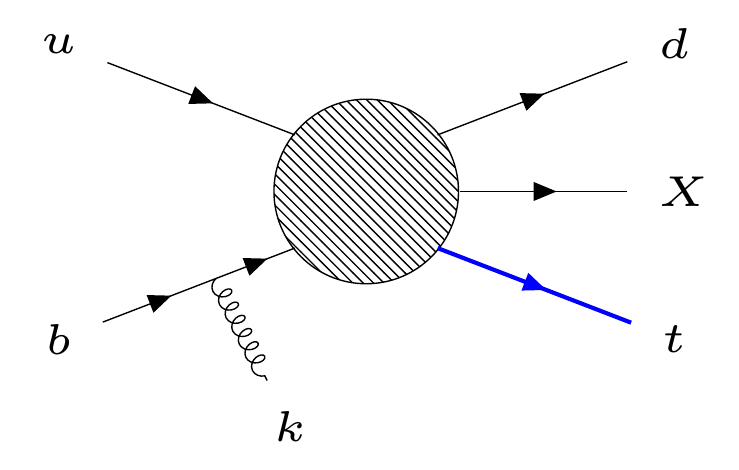}
    \end{subfigure}
        \begin{subfigure}[t]{0.49\textwidth}
        \centering
        \vskip 0pt 
        \includegraphics{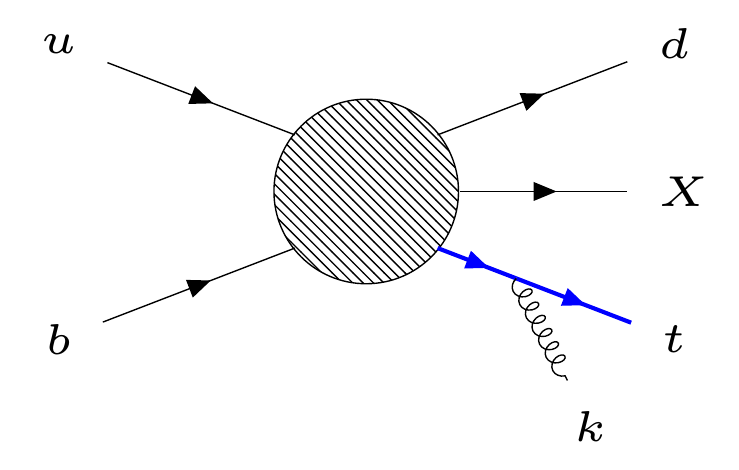}
    \end{subfigure}
        \caption{Leading order and the relevant real emission contributions to a single top production process. The blob in the center represents the function $\hatN$. We emphasise that there is no colour transfer from the light quark line to the heavy quark
        line, see text for details.}
        \label{fig1}
\end{figure}

The gluon can be emitted from the ``light'' quark line (i.e. the fermion line going from the up to the down quark)
or from the  ``heavy'' quark line (i.e. the one from the bottom quark to the top quark). However, 
since the process is mediated by an exchange of a colourless $W$-boson, the two contributions do not interfere
because of colour conservation.  As explained in ref.~\cite{Caola:2021kzt},
emissions off the light-quark line cannot produce linear power corrections; for this reason we do not discuss them
further and focus instead  on the emissions off the heavy quark line. 

It is also explained in ref.~\cite{Caola:2021kzt} that one can only obtain ${\cal O}(\lambda)$ contributions
to  the cross section of the process eq.~(\ref{eq8.1}) if the gluon $g(k)$ is soft. However, since
the leading term in the soft expansion corresponds to ${\cal O}(\lambda^0)$,  the \emph{first sub-leading}
term in the soft expansion is required. Such term can be obtained in a process-independent way using
the LBK theorem \cite{Low:1958sn,Burnett:1967km}, as we now explain. 

We write the amplitude extracting the strong coupling constant, the colour factor and the gluon polarisation vector. It reads 
\begin{equation}{\cal A}_{\rm real} = \gs T^{a}_{ij} \epsilon_\mu {\cal M}^\mu,
\end{equation}
where $a,i,j$ are the gluon, top-quark and $b$-quark colour indices and $\epsilon$ is the gluon
polarisation vector.   The reduced amplitude ${\cal M}^\mu$ reads 
\begin{equation}
\begin{split} 
  \mathcal{M}^\mu & =
  \bar u(\qtop) \gamma^\mu \frac{\hatqtop + \hatk + \mt }{\dtop} \hatN(\qtop+k,\pb,\qd,...) u(\pb)
  \\
&  +
\bar u(\qtop) \hatN(\qtop,\pb-k,\qd,...) \frac{\hatpb - \hatk}{d_b} \gamma^\mu u(\pb)
+ {\cal M}^\mu_{\rm reg}(\qtop,\pb,\qd,..| k),
\label{eq8.3}
\end{split} 
\end{equation}
where $\dtop = (\qtop+k)^2 - \mt^2 = 2\qtop k + \lambda^2$ and $d_b = (\pb - k)^2 = -2\pb k + \lambda^2$.
The three terms on the right-hand side of eq.~(\ref{eq8.3})
describe
contributions where a gluon is emitted off  an external
top-quark line, an external $b$-quark line and, finally, off any internal part of the ``heavy'' line of the process,
respectively. They are illustrated in Fig.~\ref{fig1}.
In the soft $k \sim \lambda \to 0$ limit, the first two terms in eq.~(\ref{eq8.3}) scale as $1/\lambda$ whereas the third
term scales as $\lambda^0$. Hence, to compute the amplitude through sub-leading terms in the soft expansion,
${\cal M}^\mu_{\rm reg}(\qtop,\pb,\qd,..| k)$ is required.

The matrix function $\hatN$, which can be  understood as a  Green's function of a Born-like process eq.~(\ref{eq1.1a})
with amputated 
$t$ and $b$ lines,  can be used to write the amplitude for the elastic no-emission process
$u(\pu) + b(\pb) \to d (\pd) + t(\ptop) + X(p_X)$
\begin{equation}
  {\cal A}_0 = \delta_{ij} \; \bar u(\ptop) \hatN (\ptop, \pb,\pd, ..) u(\pb).
  \label{eq8.3a}
\end{equation}
We note that we always assume  that the energy-momentum conservation condition has been used to express the 
function $\hatN$ in eqs.~(\ref{eq8.3},\;\ref{eq8.3a}) through a unique set of momenta.

In general, diagrams where gluons are only emitted from the external
$t$ and $b$ legs are not gauge invariant on their own; this fact can be used
to determine the amplitude  ${\cal M}^\mu_{\rm reg}(\qtop,\pb,\qd,..| k)$ \cite{Low:1958sn,Burnett:1967km}.
To this end,  we compute the  scalar product of  ${\cal M}^\mu$ with $k_\mu$ and  demand
that the result vanishes, as required by current conservation.  We then find 
\begin{equation}
0 = \bar u_t \hatN(\qtop+k,\pb,\qd,...) u_b - \bar u_t \hatN(\qtop,\pb-k,\qd,...)  u_b
+ k_\mu {\cal M}^\mu_{\rm reg}(\qtop,\pb,\qd,\dots | k),
\label{eq8.6}
\end{equation}
where, for ease of notation,  we do not display the arguments of the external spinors, i.e.  
$ \bar u_t(\qtop) \Rightarrow \bar u_t $ and $u_b(\pb) \Rightarrow u_b $.
We will employ this notation through the end of this section.

We solve eq.~(\ref{eq8.6}) to zeroth order in the gluon momentum $k$ by expanding the function $\hatN$
and the function ${\cal M}^\mu_{\rm reg}$  in Taylor series  in $k$.
Neglecting terms of order $k^2$, we find 
\begin{equation}
0 = k^\mu \bar u_t \left [ \frac{  \partial \hatN(\qtop,\pb,\qd,...)}{ \partial \qtop^\mu} +  \frac{\partial
  \hatN(\qtop,\pb,\qd,...)}{\partial \pb^\mu}  \right ]  u_b
+ k_\mu {\cal M}^\mu_{\rm ext}(\qtop,\pb,\qd,\dots | k = 0).
\end{equation}
This equation should hold for any $k$; therefore 
\begin{equation}\label{eq:Mmu}
   {\cal M}^\mu_{\rm ext}(\qtop,\pb,\qd,...|k = 0)
   = -\bar u_t \left [ \frac{  \partial \hatN(\qtop,\pb,\qd,...)}{ \partial q_{t, \mu}} +  \frac{\partial
  \hatN(\qtop,\pb,\qd,...)}{\partial p_{b, \mu}}  \right ]  u_b.
\end{equation}

To proceed further, we simplify the expressions for diagrams where the gluon is  emitted off the external lines.
We write
\begin{equation}
\bar u_t \gamma^\mu \frac{\hatqtop + \hatk + \mt}{\dtop}
=\bar u_t \frac{2 \qtop^\mu + k^\mu + \sigma^{\mu \nu} k_\nu}{\dtop}
 = \bar u_t \left [ J_t^\mu + \hatS_t^\mu \right ],
\end{equation}
where $\sigma^{\mu \nu} = \frac{1}{2} [\gamma^\mu, \gamma^\nu]$ and we introduced spin-independent and
spin-dependent currents
\begin{equation}
J_t^\mu = \frac{2\qtop + k^\mu}{\dtop},\;\;\;\; \hatS_t^\mu = \frac{\sigma^{\mu \nu} k_\nu}{\dtop},
\end{equation}
which describe the gluon emission off the top quark.   Similarly,
\begin{equation}
\frac{\hatpb - \hatk}{d_b} \gamma^\mu u_b
= \frac{2 \pb^\mu  - k^\mu + \sigma^{\mu \nu} k_\nu}{d_b}  u_b  =
 \left [J_b^\mu + \hatS_b^\mu \right ] u_b, 
\end{equation}
where
\begin{equation}
J_b^\mu = \frac{2 \pb^\mu  - k^\mu}{d_b},\;\;\;\; \hatS_b^\mu = \frac{\sigma^{\mu \nu} k_\nu}{d_b}.
\end{equation}

We can use these results to write the amplitude for a single gluon emission through the first sub-leading
terms in the soft expansion. We find 
\begin{equation}
\begin{split} 
   {\cal M}^\mu & =
   J_t^\mu \bar u_t \hatN(\qtop+k,\pb,\qd,..) u_b + J_b^\mu \bar u_t \hatN(\qtop,\pb-k,\qd,..) u_b
   \\
   & + \bar u_t  \left [ \hatS_t^{\mu} \hatN(\qtop,\pb,\qd,..) +\hatN(\qtop,\pb,\qd,..) \hatS_b^\mu \right ] u_b
   \\
   &
      - \bar u_t \left [ \frac{  \partial \hatN(\qtop,\pb,\qd,...)}{ \partial q_{t, \mu}}
   +  \frac{\partial    \hatN(\qtop,\pb,\qd,...)}{\partial p_{b, \mu}}  \right ]  u_b.
   \end{split}
\label{eq8.14}
\end{equation}

We  can further simplify this expression by expanding the first two terms to first sub-leading
order in $k$ and combining them with the last two terms in the above formula.  We find
\begin{equation}
\begin{split}
   {\cal M}^\mu & =
   J^\mu \bar u_t \hatN(\qtop,\pb,\qd,..) u_b 
   +   \bar u_t ( L^\mu\hatN(\qtop,\pb,\qd,..) ) u_b
   \\
   & + \bar u_t  \left [ \hatS_t^{\mu} \hatN(\qtop,\pb,\qd,..)
     +\hatN(\qtop,\pb,\qd,..) \hatS_b^\mu \right ] u_b.
\end{split}
\label{eq8.15}
\end{equation}
In writing eq.~(\ref{eq8.15}) we introduced the notation 
\begin{equation}
J^\mu = J_t^\mu + J_b^\mu,\;\;\;\;L^\mu = L_t^\mu - L_b^\mu,
\end{equation}
with 
\begin{equation}
L_t^\mu = J_t^\mu k^\nu \frac{\partial }{\partial \qtop^\nu} - \frac{\partial }{\partial q_{t, \mu}}
\end{equation}
and
\begin{equation}
L_b^\mu = J_b^\mu k^\nu \frac{\partial }{\partial \pb^\nu} +  \frac{\partial }{\partial p_{b, \mu}}.
\end{equation}

Eq.~(\ref{eq8.15}) gives the desired result as it expresses the amplitude that describes the  emission of a
single soft gluon  through an elastic amplitude and its derivatives.
 Further  simplifications occur if we square the amplitude and sum over the polarisations of the external
particles.  To see this, we write the conjugate amplitude
\begin{equation}
  {\cal M}^{\mu,+} =
  J^\mu \bar u_b {\barhatN}\defargs u_t
  +   \bar u_b ( L^\mu {\barhatN}\defargs ) u_t
  - \bar u_b  \left [  {\barhatN}\defargs \hatS_t^{\mu}
    + \hatS_b^\mu {\barhatN}\defargs  \right ] u_t ,
\label{eq8.16}
\end{equation}
(where for ease of notation we have dropped the arguments of $\hatN$),
and use it to compute  the squared amplitude summed over  polarisations of the external
particles through the first sub-leading term in the soft expansion. We obtain 
\begin{align}
& |{\cal M}|^2 = -g_{\mu \nu} {\cal M}^\mu {\cal M}^{\nu,+}
  = -J^\mu J_\mu  F_{\rm LO}(\qtop,\pb,\qd,...) \nonumber 
        \\
& - J_\mu {\rm Tr} \left [ ( \hatqtop + \mt) \hatN \hatpb L^\mu {\barhatN}\defargs  \right ]
-J_\mu  {\rm Tr} \left [ ( \hatqtop + \mt) ( L^\mu \hatN ) 
  \hatpb  {\barhatN}\defargs  \right ]
\\
&
+J_\mu {\rm Tr} \left [ [\hatS^\mu_t,\hatqtop] \hatN \hatpb {\barhatN} \right ]
+J_\mu {\rm Tr} \left [ ( \hatqtop + \mt)  \hatN [\hatpb,\hatS^\mu_b ] {\barhatN} \right ],
\nonumber 
\end{align} 
where
\begin{equation}
F_{\rm LO}(\qtop,\pb,\qd,...) = {\rm Tr} \left [ ( \hatqtop + \mt) \hatN \hatpb {\barhatN}\defargs  \right ].
\end{equation}

Since
\begin{equation}
   [\hatS_t^\mu,\hatqtop] = -L_t^\mu \hatqtop = -L^\mu \hatqtop,
   \;\;\;\;\; [\hatpb, \hatS_b^\mu] = L_b^\mu \hatpb = - L^\mu \hatpb,
\end{equation}
we find
\begin{align}
  &  |{\cal M}|^2   = -J^\mu J_\mu F_{\rm LO}(\qtop,\pb,\qd,...)
\nonumber 
  \\
& - J_\mu {\rm Tr} \left [ ( \hatqtop + \mt) \hatN \hatpb L^\mu {\barhatN}\defargs  \right ]
-J_\mu  {\rm Tr} \left [ ( \hatqtop + \mt) ( L^\mu \hatN ) 
  \hatpb  {\barhatN}\defargs  \right ]
\\
&
-J_\mu {\rm Tr} \left [ ( L^\mu (\hatqtop + \mt))  \hatN \hatpb {\barhatN} \right ]
-J_\mu {\rm Tr} \left [ ( \hatqtop + \mt)  \hatN (L^\mu \hatpb) {\barhatN} \right ].
\nonumber 
\end{align}

Making use of the fact that $L_\mu$ is a linear differential operator, we combine
the last four terms to obtain a derivative 
of the leading order function $F_{\rm LO}$. The final result reads 
\begin{align}
& |{\cal M}|^2 
= -J^\mu J_\mu F_{\rm LO}(\qtop,\pb,\qd,...)
-J_\mu L^\mu F_{\rm LO}(\qtop,\pb,\qd,...).
\label{eq2.22}
\end{align}

In order to obtain the ${\cal O}(\lambda)$ contribution to the cross
section of a generic single top production process due to the real gluon
emission, we need to integrate eq.~(\ref{eq2.22}) over the phase space
of the final state particles.  It was pointed out in
ref.~\cite{Caola:2021kzt} that the relevant integration can be
performed in a process-independent manner provided that an approximate
momentum mapping, that factorises integration over the gluon momentum,
is performed.

To construct such a mapping,   we redefine the momenta of the top quark and
of the outgoing massless quark   as follows
\begin{equation}
\begin{split}
& \qtop = \ptop - k + \frac{\ptop   k}{ \ptop    \pd}  \pd,\\
& \qd = \pd - \frac{ \ptop   k}{ \ptop     \pd}  \pd.
\end{split} \label{eq:momentamapping}
\end{equation}
We note that through ${\cal O}(k^2)$, $\qtop^2 = \ptop^2  = \mt^2 $ and $\qd^2 = \pd^2 = 0$. 
Furthermore, when written in terms of $\ptop$ and $\pd$, the final state four-momentum loses
its dependence on the gluon momentum $k$ 
\begin{equation}
\qtop + \qd + k + p_X  = \ptop + \pd + p_X. 
\end{equation}
The Jacobians of the respective transformations  read 
\begin{equation}
\begin{split} 
  & {\rm det} \left | \frac{ \partial \qtop^\mu}{\partial \ptop^\nu} \right | = 1 + \frac{k   \pd }{\ptop   \pd}+{\cal O}(k^2),
  \\
  & {\rm det} \left | \frac{ \partial \qd^\mu}{\partial \pd^\nu} \right |
  = 1 -3  \frac{k   \ptop }{ \ptop   \pd} + {\cal O}(k^2).
\end{split} 
\end{equation}
Also, we find
\begin{equation}
\delta(\qd^2) = \delta( \pd^2) \left (1 + 2 \frac{ \ptop   k}{ \ptop   \pd} + {\cal O}(k^2) \right ).
\end{equation}
The above formulae can be used to re-write the partonic phase space as follows 
\begin{equation}
\begin{split}
  & {\rm d Lips}(\pu,\pb;\qd,\qtop ,p_X,k)
  \\
  &   = {\rm d Lips}(\pu,\pb; \pd, \ptop,p_X)
   \frac{{\rm d}^4 k}{(2\pi)^3} \delta_+(k^2 - \lambda^2) \times \left [
     1 + \frac{k   \pd }{ \ptop   \pd} - \frac{ \ptop k}{ \ptop   \pd} \right ]
   +{\cal O}(k^2).
   \label{eq2.27}
   \end{split}
\end{equation}
We note that the above expression should also include an  upper bound on the integration over the
momentum $k$ which depends on the other momenta.
However, such a bound plays no role for the extraction of ${\cal O}(\lambda)$
contributions which arise exclusively from the low integration boundary for the momentum $k$. 

The rest of the calculation is straightforward. We use eq.~(\ref{eq2.27})  for the phase space together with
the expression for the matrix element squared given in eq.~(\ref{eq2.22}). We then use the momenta mapping
of eq.~(\ref{eq:momentamapping})
in eq.~(\ref{eq2.22}), expand the matrix element squared through the first sub-leading terms in $k$ and integrate
over $k$ to extract  the ${\cal O}(\lambda)$ terms.

Although the above procedure is straightforward, we point out   that care is required when expanding the inverse
top propagator $\dtop = 2 \qtop k + \lambda^2$ since it also has to be expressed through $\ptop$, expanded in $k \sim \lambda$
and then integrated.  For $\dtop$ we obtain 
\begin{equation}
\dtop  = 2 \qtop k + k^2 = 2 \ptop k  - k^2 + 2\frac{ ( \ptop   k) ( \pd   k)}{ \ptop   \pd},
\end{equation}
and
\begin{equation}
 \frac{1}{\dtop}
 = \frac{1}{2 \ptop    k} \left( 1 + \frac{k^2}{2 \ptop   k}
   -  \frac{  \pd   k }{ \ptop   \pd } +{\cal O}{(k^2)} \right).
\end{equation}
On the contrary,   the expansion of $1/\db$ is simple since the momentum $\pb$ is not subject to momentum mapping. 

Upon combining the approximate expressions for the matrix element squared and the phase space, 
the dependence on the gluon momentum becomes explicit through the required order in
the soft expansion.  The corresponding integrals over $k$ are given in Appendix~\ref{app:integrals}.
Finally, putting everything together,
we find the following result for the 
${\cal O}(\lambda)$ correction to the real-emission contribution to the differential cross section\footnote{
  The  operator ${\cal T}_\lambda$ which appears in eq.~(\ref{eq:r}) 
extracts the ${\cal O}(\lambda)$ contribution from a quantity it acts upon.} 
\begin{equation}
  \begin{split} 
    & {\cal T}_\lambda  \left[ \sigma_{\rm R} \right ]=
    \frac{\alpha_s \Cf}{2 \pi }\frac{\pi \lambda}{\mt}
    \int  {\rm d Lips}(\pu,\pb; \pd, \ptop,p_X)
    \Bigg [ 
    \left ( \frac{3}{2}  - \frac{\mt^2}{ \pd  \ptop} - \frac{\mt^2}{  \ptop \pb}  \right ) 
    \\
    &   - \frac{ \mt^2}{  \pd  \ptop } \pd^\mu  \left ( \frac{\partial }{\partial \pd^\mu } - \frac{\partial }{\partial \ptop^\mu}
    \right )  -  \frac{ \mt^2}{ \ptop \pb} \pb^\mu \left ( \frac{\partial }{\partial \pb^\mu} + \frac{\partial}{\partial \ptop^\mu} \right )
    \Bigg  ] F_{\rm LO}.
    \label{eq:r}
\end{split}
\end{equation}
As we will see later, we do not need to compute the derivatives of the leading order amplitude squared explicitly
because, as it turns out, all such terms get cancelled once the virtual
corrections and the  renormalisation terms
are added to the real emission contribution.

\section{Virtual corrections}
\label{sect:virt}

Similar to the case of the real emission corrections discussed in the previous section, the 
${\cal O}(\lambda)$ contributions to the  virtual corrections can only arise from the region of
soft $k \sim \lambda$ loop momenta. Our goal, therefore, is to construct the soft expansion
of the one-loop virtual corrections
to the generic single top production processes $u(\pu) + b(\pb) \to d(\pd) + t(\ptop) + X$.
We focus on the corrections to the ``heavy'' quark
line and we remind the reader that, thanks to colour conservation,  one-loop diagrams where gluons are exchanged
between ``light'' and ``heavy'' quark lines do not contribute to the cross section at this perturbative order.

\begin{figure}
        \begin{subfigure}[t]{0.49\textwidth}
        \centering
        \vskip 0pt
        \includegraphics{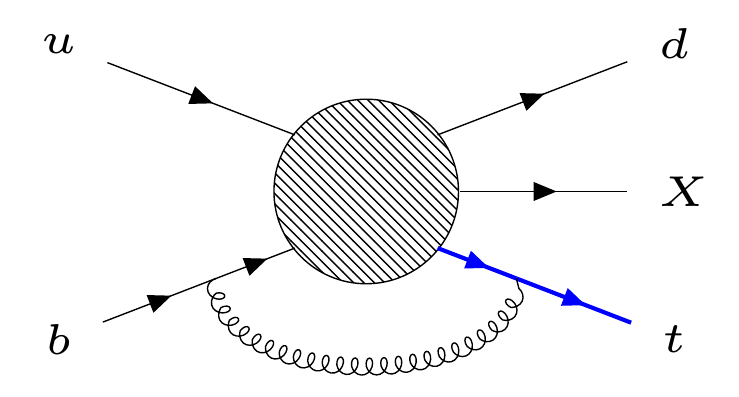}
    \end{subfigure}
        \begin{subfigure}[t]{0.49\textwidth}
        \centering
        \vskip 0pt
        \includegraphics{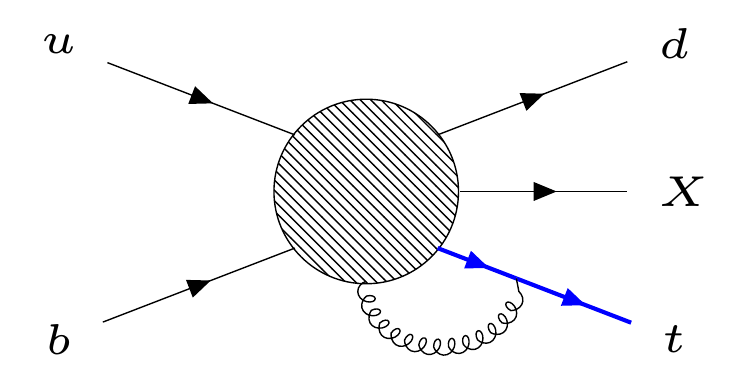}
        \end{subfigure}
                \begin{subfigure}[t]{0.49\textwidth}
        \centering
        \vskip 0pt
        \includegraphics{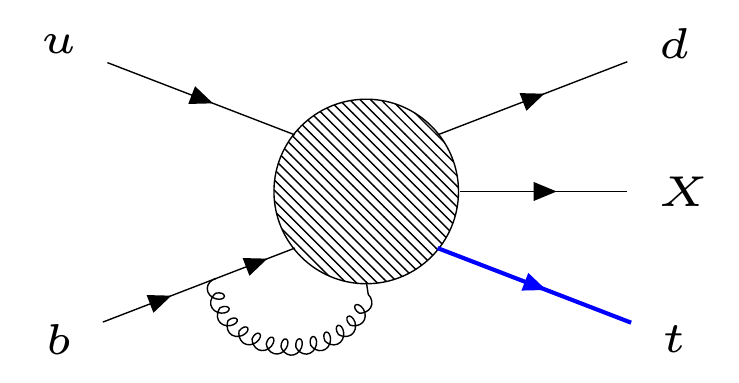}
    \end{subfigure}
    \begin{subfigure}[t]{0.49\textwidth}
        \centering
        \vskip 0pt
        \includegraphics{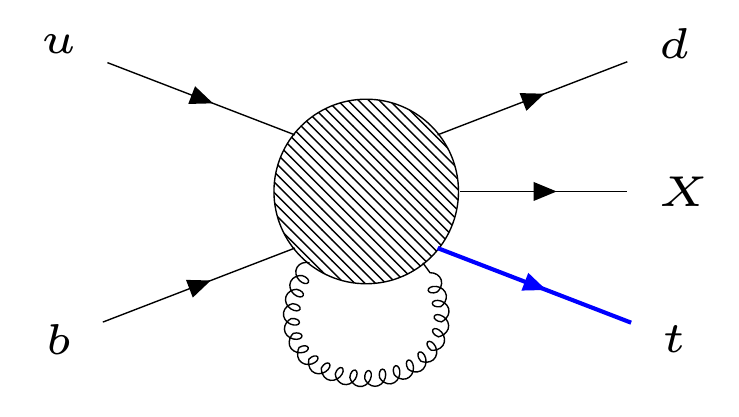}
    \end{subfigure}
    \caption{Loop contributions to single top production that  need to be considered.
We emphasise that there is no colour transfer from the light quark line to the heavy quark
        line, see text for details.}
        \label{fig2}
\end{figure}

We write
\begin{equation}
{\cal A}_{\rm virt} = \gs^2 \Cf \delta_{ij} {\cal M}_{\rm virt},
\end{equation}
where $i,j$ are the colour indices of the top quark and the bottom quark. We note that  the
one-loop corrections to the ``heavy'' line can be written as the sum of four contributions (see Fig.~\ref{fig2})
\begin{equation}
   {\cal M}_{\rm virt} = \sum \limits_{i \in \{a,b,c,d\} }{\cal M}^{(i)}_{\rm virt},
\end{equation}
where
\begin{equation}
\begin{split} 
& {\cal M}_{\rm virt}^{(a)}
   =  \int \frac{{\rm d}^4 k}{(2 \pi)^4} \frac{-i}{k^2 - \lambda^2}
   \left [ \bar u_t \gamma^{\alpha} \frac{( \hatptop + \hatk + \mt)}{\dtop}
     \hatN (\ptop+k,\pb + k,..) \frac{(\hatpb+\hatk)}{d_b} \gamma_\alpha u_b \right ],
     \\
     & {\cal M}_{\rm virt}^{(b)}
   =  \int \frac{{\rm d}^4 k}{(2 \pi)^4} \frac{-i}{k^2 - \lambda^2}
   \left [ \bar u_t \gamma_{\alpha} \frac{( \hatptop + \hatk + \mt)}{\dtop}
     \hatN_{1 g}^{\alpha}  (\ptop+k,\pb ,..,|-k)  u_b \right ],
     \\
     & {\cal M}_{\rm virt}^{(c)}
   = \int \frac{{\rm d}^4 k}{(2 \pi)^4} \frac{-i}{k^2 - \lambda^2}
     \left [ \bar u_t 
       \hatN^{\alpha}_{1g}  (\ptop ,\pb + k,..,|k) \frac{(\hatpb+\hatk)}{d_b} \gamma_\alpha u_b \right ],
       \\
     & {\cal M}_{\rm virt}^{(d)}
   =  \int \frac{{\rm d}^4 k}{(2 \pi)^4} \frac{-i g_{\alpha \beta} }{k^2 - \lambda^2} \left [ 
   \bar u_t 
     \hatN_{2g}^{\alpha \beta} (\ptop,\pb,..|k,-k)  u_b \right ].
\end{split} 
\end{equation}
By a slight abuse of notation,  we use $d_b = (\pb + k)^2$ in this section,
and we continue to denote the external spinors as   $\bar u_t = \bar u_t(\ptop)$ and $u_b =  u_b(\pb)$. 
The quantities $\hatN (\ptop+k,\pb + k,..)$, $ \hatN_{1g}^{\alpha}  (\ptop+k,\pb ,..,|-k)$ and
      $ \hatN_{2 g}^{\alpha}  (\ptop+k,\pb ,..,|k,-k)$ are functions  that contribute to  processes  where the corresponding
number of gluons\footnote{Of course, these ``gluons'' are no different from photons since
  no non-Abelian interactions need to be considered.} (from zero to two)
are emitted.  We note that these functions do not include contributions where 
      gluons  are emitted from the external ($t$ and $b$) legs; for this reason all  of them   have   smooth $k \to 0$
      limits.  To compute the ${\cal O}(\lambda)$ contribution to the differential cross section only the $k \sim \lambda$
      integration region is relevant;  as a result, 
      all these functions can be expanded in  Taylor series at small $k$.

      A simple power counting suggests
      that ${\cal M}_{\rm virt}^{(d)}$ cannot provide an ${\cal O}(\lambda)$ contribution  and therefore  can be neglected, the function
      $\hatN_{1g}$ is needed at $k = 0$ and the function $\hatN$ is needed through  linear terms in $k$.
      Hence, we can write 
      \begin{equation}
      \hatN(\ptop+k,\pb+k,...) = \hatN(\ptop,\pb,...) + k_\mu D_p^\mu \hatN(\ptop,\pb,...) + {\cal O}(k^2),
      \label{eq3.4}
      \end{equation}
      where
      \begin{equation}
D_p^\mu = \frac{\partial }{\partial p_{t,\mu}} + \frac{\partial }{\partial p_{b,\mu}}.
\end{equation}

The function $\hatN_{1g}$  needs to be known at $k=0$. Following  the discussion of the real emission contribution (cf. eq.~(\ref{eq:Mmu})),
we find 
\begin{equation}
\bar u_t \hatN_{1g}^{\alpha}(\ptop,\pb,..|k = 0) u_b = - \bar u_t D_p^\alpha \hatN u_b.
\label{eq3.6}
\end{equation}

Eqs.~(\ref{eq3.4},\;\ref{eq3.6}) are  sufficient to write an approximate expression for the virtual corrections.  The manipulations
are nearly identical to what has been discussed in the context of the real emission contribution  in the previous
section. We obtain 
\begin{equation}
\bar u_t \gamma^{\alpha} \frac{( \hatptop + \hatk + \mt)}{\dtop} = \bar u_t \left ( J_t^\alpha + \hatS_t^{\alpha} \right ),
\end{equation}
and
\begin{equation}
\frac{( \hatpb + \hatk)}{d_b} \gamma^\alpha u_b   = \left ( J_b^\alpha - \hatS_b^{\alpha} \right ) u_b,
\end{equation}
where
\begin{equation}
J_t^\alpha = \frac{2 \ptop^\alpha + k^\alpha}{\dtop},\;\;\;\hatS_t^\alpha = \frac{\sigma^{\alpha \beta} k_\beta}{\dtop},
\end{equation}
and
\begin{equation}
J_b^\alpha = \frac{2 \pb^\alpha + k^\alpha}{d_b},\;\;\;\hatS_b^\alpha = \frac{\sigma^{\alpha \beta} k_\beta}{d_b}.
\end{equation}

Using these expressions and keeping only those terms that can provide linear power corrections,
we find 
\begin{equation}
\begin{split}
  & {\cal M}_{\rm virt} = \int \frac{{\rm d}^4 k}{(2 \pi)^4} \frac{-i}{k^2 - \lambda^2}
  \Bigg [ J_t^\alpha J_{b,\alpha} \; \bar u_t \left (  \hatN(\ptop,\pb,..) + k^\mu D_{p,\mu} \hatN(\ptop,\pb,..)\right ) u_b
     \\
&    - J_t^\alpha \bar u_t  \hatN(\ptop,\pb,..) \hatS_{b,\alpha} u_b
    + J_b^\alpha \bar u_t \hatS_{t,\alpha}  \hatN(\ptop,\pb,..) u_b
   -(J_t^\alpha +J_b^\alpha) \bar u_t D_{p,\alpha} \hatN u_b 
    \Bigg ].
  \end{split}
\end{equation}
Similar to the case of the real emission corrections, the dependence on the loop momentum has been made explicit so that
the integration over $k$ becomes possible.  However, it is beneficial to compute the correction to the matrix element
squared before integrating over $k$. We find
\begin{equation}
\begin{split}
& \delta_{\rm virt}[ {\cal M} {\cal M}^+ ]
= \int \frac{{\rm d}^4 k}{(2 \pi)^4} \frac{-i}{k^2 - \lambda^2}
\Bigg [
  2 J_t^\alpha J_{b,\alpha} F_{\rm LO}
  \\
&   + J_t^\alpha J_{b,\alpha} k^\mu \; {\rm Tr} \left [ (\hatptop+\mt) (D_{p,\mu} \hatN) \hatpb \bar {\hatN} +
    (\hatptop+\mt) \hatN \hatpb (D_{p,\mu} )  {\bar {\hatN}} \right ]
  \\
  & -(J_t^\alpha + J_b^\alpha) {\rm Tr} \left [ (\hatptop + \mt) ( D_{p,\alpha}\hatN ) \hatpb \barhatN
        + (\hatptop + \mt) \hatN \hatpb ( D_{p,\alpha} \barhatN )
    \right ]
\\
& +  J_b^\alpha {\rm Tr} \left[ [\hatptop,\hatS_{t,\alpha} ]\hatN \hatpb \barhatN \right ]
- J_t^\alpha {\rm Tr} \left[ (\hatptop + \mt) \hatN [\hatS_{b, \alpha},\hatpb] \barhatN \right ]
\Bigg  ],
\end{split}
\end{equation}
where ${\cal M}={\cal M}_0+{\cal M}_{\rm virt}$.
We can further simplify this expression  following the steps already discussed in the context
of the real emission contribution.  Indeed, using 
\begin{equation}
\begin{split} 
  &  [\hatptop, \hatS_t^\alpha] = \left ( J_t^\alpha k^\nu \frac{\partial }{\partial p_{t}^{\nu} } - \frac{\partial }{\partial p_{t,\alpha} } \right )
  \hatptop =    L_t^\alpha \hatptop,
  \\
    &  [\hatpb, \hatS_b^\alpha] =  \left ( J_b^\alpha k^\nu \frac{\partial }{\partial p_{b}^{\nu} } - \frac{\partial }{\partial p_{b,\alpha} } \right )
  \hatpb =    L_b^\alpha \hatpb,
\end{split}
    \end{equation}
we arrive at 
    \begin{equation}
\begin{split}
& \delta [ {\cal M}{\cal M}^+ ]_{\rm virt}
= \int \frac{{\rm d}^4 k}{(2 \pi)^4} \frac{-i}{k^2 - \lambda^2}
\Bigg [
  2 J_t^\alpha J_{b,\alpha} F_{\rm LO}
  \\
&   + J_t^\alpha J_{b,\alpha} k^\mu \; {\rm Tr} \left [ (\hatptop+\mt) (D_{p,\mu}\hatN) \hatpb \barhatN +
    (\hatptop+\mt)\hatN \hatpb (D_{p,\mu}   {\barhatN}) \right ]
  \\
  & -(J_t^\alpha + J_b^\alpha) {\rm Tr} \left [ (\hatptop + \mt) ( D_{p,\alpha}\hatN ) \hatpb \barhatN
        + (\hatptop + \mt) \hatN \hatpb ( D_{p,\alpha} \barhatN )
    \right ]
\\
& +  J_b^\alpha {\rm Tr} \left[ (  L_{t,\alpha} \hatptop )\hatN  \hatpb \barhatN \right ]
 + J_t^\alpha {\rm Tr} \left[ (\hatptop + \mt) \hatN ( L_{b,\alpha} \hatpb)  \barhatN \right ]
 \Bigg  ].
\label{eq10.14}
\end{split}
\end{equation}
To simplify this expression further, we take the  terms $ J_{(t,b)}^{\alpha} k^{\mu} \partial /\partial p_{t,b}^\mu $
from $L_{t,b}^\alpha$  and combine them  with  the similar terms  in the second line of eq.~(\ref{eq10.14}). 
We finally obtain 
\begin{equation}
\begin{split} 
& \delta [ {\cal M} {\cal M}^+ ]_{\rm virt}
= \int \frac{{\rm d}^4 k}{(2 \pi)^4} \frac{-i}{k^2 - \lambda^2}
\Bigg [ 2 J_t^\alpha J_{b,\alpha} F_{\rm LO}
  \\
  & + J_t^\alpha J_{b,\alpha} k^\mu \; D_{p,\mu} F_{\rm LO} -(J_t^\alpha + J_b^\alpha) D_{p,\alpha} F_{\rm LO}
      \\
& + J_t^\alpha {\rm Tr} \left[ ( D_{p,\alpha}  \hatptop )\hatN  \hatpb \barhatN \right ]
 + J_b^\alpha {\rm Tr} \left[ (\hatptop + \mt) \hatN ( D_{p,\alpha} \hatpb)  \barhatN \right ]
  \Bigg ].
\end{split}
\end{equation}

The loop momentum $k$ in the above expression is contained in 
the currents $J_{t,b}^\mu$ and also appears explicitly in a few terms. Hence, 
it becomes possible to integrate over $k$.  The needed integrals are given in  Appendix~\ref{app:integrals}.
Finally, putting everything together, we obtain
\begin{equation}
\begin{split}
  {\cal T}_\lambda \left [  \sigma_V  \right ]
  =
 & 
     -\frac{\as \Cf}{2 \pi} \frac{\pi \lambda}{\mt} \int  {\rm d Lips}_{\rm LO}
     \Bigg [  {\rm Tr} \left [ \hatptop \hatN \hatpb {\bar  \hatN}  \right ]
       \\
       &
       +\left ( \frac{2 \ptop \pb - \mt^2}{ \ptop\pb} - \frac{\mt^2}{\ptop \pb} \pb^\mu D_{p,\mu} \right ) F_{\rm LO}
       \Bigg ],
     \label{eq:virtcorrs}
\end{split}
\end{equation}
where we have introduced the notation $ {\rm d Lips}_{\rm LO}={\rm d Lips}(\pu,\pb;\pd,\ptop,p_X)$.

\section{Renormalisation contributions}
\label{sect:renorm}

The above result for the virtual corrections has  to be supplemented with the renormalisation
contributions. Two of them (the wave function renormalisation of the
external top quark and the top quark mass counter-term in the pole-mass scheme) provide 
${\cal O}(\lambda)$ corrections to the cross section.

The two renormalisation constants can be computed using standard methods and read
\begin{equation}
\begin{split} 
  & Z_m = 1 + \frac{ \Cf \gs^2 \mt^{-2\ep} \Gamma(1+\ep) }{(4\pi)^{d/2} } \left [ -\frac{3}{\ep} - 4 + \frac{ 2  \pi \lambda}{\mt} +{\cal O}
    \left(\frac{\lambda^2}{\mt^2}\right)  \right ],
  \\
  & Z_2 = 1 + \frac{ \Cf \gs^2 \mt^{-2\ep} \Gamma(1+\ep) }{(4\pi)^{d/2} } \left [ -\frac{1}{\ep} - 4
    + 4 \ln \frac{\mt}{\lambda} + \frac{ 3 \lambda \pi}{\mt} +{\cal O}
    \left(\frac{\lambda^2}{\mt^2}\right) \right ].  
\end{split} 
  \end{equation}
For the purpose of our discussion only the ${\cal O}(\lambda)$ contributions to $Z_2$ and $Z_m$ are relevant.

It is straightforward to add the wave function renormalisation contribution to the virtual corrections.
The mass counter-term, on the other hand, is only relevant for the internal top quark lines.
Since the relation between the bare mass  $m_0$ and the pole mass $\mt$ is given by
$m_0 = Z_m \mt$,  we find
\begin{equation}
\frac{1}{\hatptop - m_{0}} =
\frac{1}{\hatptop - \mt -(Z_m-1)\mt}
\approx
\frac{1}{\hatptop - \mt}  + (Z_m-1)\mt\frac{\partial}{\partial \mt}\frac{1}{\hatptop - \mt} 
\end{equation}
where
\begin{equation}
  {\cal T}_\lambda \left [ (Z_m - 1 ) \right ] \mt
  = \frac{\Cf \alpha_s}{2 \pi} \pi \lambda.
\end{equation}

Putting everything together,  we find the following result for the
renormalisation contributions to the   cross section 
\begin{equation}
  \begin{split} 
    {\cal T}_\lambda \left [  \sigma_{\rm ren}  \right ]
     = 
    \frac{\alpha_s \Cf}{2 \pi }\frac{\pi \lambda}{\mt}
    \int  {\rm d Lips}_{\rm LO}
    \Bigg  [  \frac{3}{2}  F_{\rm LO}
    & +
    \mt {\rm Tr} \left [ ( \hatptop + \mt) \frac{ \partial\hatN}{\partial \mt} \hatpb \barhatN \right ]
    \\
    &
    +
    \mt {\rm Tr} \left [ ( \hatptop + \mt)\hatN  \hatpb \frac{ \partial \barhatN}{\partial \mt}  \right ]
   \Bigg  ].
  \end{split} \label{eq:rencontr}
\end{equation}

\section{Redefining the mass}
\label{sect:mass}
        
It is well known that the use of the pole quark mass in physical predictions is one of the sources of  linear power
corrections.  Such corrections are artificial and can be removed by employing   one of the many
short-distance mass schemes  \cite{tHooft:1973mfk,Czarnecki:1997sz,Beneke:1998rk,Hoang:1999ye,Hoang:2008yj}  instead;
we will refer to  masses in such schemes  as $\tilde m_t$. 
Hence, we need to derive a formula that provides  a change in the
cross section due to the change of the top quark mass.

To do this, it  is important to recognise that such a  dependence arises for two distinct reasons:
1) the \emph{implicit} dependence  of the energies of  the final state particles on $\mt$ and 2) the \emph{explicit}
dependence  of the matrix element squared on this parameter.

The explicit dependence is computed by writing  $\mt = \tilde{m}_t + \delta \mt$ in the function $F_{\rm LO}$.
The corresponding change in the leading order  cross section reads 
\begin{equation}
\begin{split} 
  \delta \sigma_{\rm mass}^{\rm expl}
  =& 
  \delta \mt \int  {\rm d Lips}_{\rm LO} \frac{\partial F_{\rm LO}  }{\partial \mt} 
  \\
  =&
  \delta \mt \int  {\rm d Lips}_{\rm LO}  \; \left ( 
   {\rm Tr} \left [ \hatone  \hatN \hatpb \barhatN \right ]
   + {\rm Tr} \left [ ( \hatptop + \mt) \left (
     \frac{ \partial\hatN}{\partial \mt} \hatpb \barhatN 
           + \hatN  \hatpb \frac{ \partial \barhatN}{\partial \mt}  \right ) \right ]
           \right ).
\end{split} 
\end{equation}

To compute the change in the cross section caused by the implicit dependence of the energies of the final state
particles on $\mt$, we redefine the momenta of the  top quark and another final state particle  that we take
to be the outgoing down quark,  and write 
        \begin{equation}
\pd = ( 1+ \kappa) \tildepd,\;\;\; \ptop = \tildeptop - \kappa \tildepd.
        \end{equation}
        It follows that 
        \begin{equation}
\ptop^2 = \mt^2 = \tildeptop^2 - 2 \kappa \tildeptop \tildepd.
        \end{equation}
        Hence, if we choose
        \begin{equation}
\kappa = -\frac{ \delta \mt^2}{2 \tilde \ptop   {\tilde p}_d},
\end{equation}
the mass-shell condition for $\tildeptop$ becomes 
\begin{equation}
\tildeptop^2 = \tildemt^2 = \mt^2 - \delta \mt^2. 
\end{equation}

Following the discussion of the momenta  mapping  of the real emission  contribution in Section~\ref{sect:real}
and
adjusting it where necessary, we find
\begin{equation}
   {\rm dLips}\left ( \pu, \pb, \pd,\ptop,p_X; \mt^2 \right ) =
   {\rm d Lips} \left (\pu,\pb,\tildepd,\tildeptop,p_X; \tildemt^2 \right ) \left ( 1+ \kappa \right ).
\end{equation}
Finally, expanding  the leading order amplitude squared we obtain the change of the cross section
due to the implicit mass change 
\begin{equation} \label{eq:implicitmass}
\begin{split} 
& \delta  \sigma^{\rm impl}_{\rm mass} = \int {\rm d Lips} \left (\pu,\pb,\tildepd,\tildeptop,p_X \right )
\left [\kappa  + \kappa \tildepd^\mu \left ( \frac{\partial }{\partial \tildepd^\mu}
  - \frac{\partial }{\partial \tildeptop^\mu}
  \right ) \right ] F_{\rm LO}(\tildeptop,\tildepd,\ldots)
\\
& = -\int {\rm d Lips} \left (\pu,\pb,\pd,\ptop,p_X \right ) \frac{\delta \mt^2}{2 \pd \ptop}
\left [ 1 +  \pd^\mu \left ( \frac{\partial }{\partial \pd^\mu}
  - \frac{\partial }{\partial \ptop^\mu}
  \right ) \right ] F_{\rm LO}(\ptop,\pd,\ldots),
\end{split}
\end{equation}
where in the last  step  we have re-labelled the momenta  $\tildeptop \Rightarrow \ptop$ and
$\tildepd \Rightarrow \pd$.
Although  short-distance masses can be defined in many different ways \cite{tHooft:1973mfk,Czarnecki:1997sz,Beneke:1998rk,Hoang:1999ye,Hoang:2008yj}, they should not
contain a linear ${\cal O}(\lambda)$ term.  Hence, for our purposes, it suffices to write 
\begin{equation}
  \mt = {\tilde m}_t  \left ( 1-   \frac{\Cf \alpha_s}{2 \pi} \frac{\pi \lambda }{\mt} \right ).
  \label{eq5.8a}
\end{equation}
It follows  that
\begin{equation}
\delta \mt = - \mt \; \frac{\Cf \alpha_s}{2 \pi} \frac{\pi \lambda }{\mt},
\;\;\;\; \delta \mt^2 = -2  \mt^2 \; \frac{\Cf \alpha_s}{2 \pi} \frac{\pi \lambda }{\mt}.
\end{equation}

Putting everything together, we finally find the change  of  the cross section due to the mass shift
\begin{equation}
\begin{split}
  \sigma_{\rm LO}(\mt) - \sigma_{\rm LO}({\tilde m}_t)
  & =  \delta  \sigma^{\rm expl}_{\rm mass} + \delta  \sigma^{\rm impl}_{\rm mass}
  = \frac{\Cf \alpha_s}{2 \pi} \frac{\pi \lambda}{\mt}
  \; \int {\rm d Lips}_{\rm LO}
  \times 
  \\
  & \Bigg [
  \frac{\mt^2}{\pd \ptop}
  \left [ 1 +  \pd^\mu \left ( \frac{\partial }{\partial \pd^\mu}
      - \frac{\partial }{\partial \ptop^\mu}
    \right ) \right ] F_{\rm LO}  -   \mt {\rm Tr} \left [ \hatone  \hatN \hatpb \barhatN \right ]
  \\
  & - \mt {\rm Tr} \left [ ( \hatptop + \mt) \left (
      \frac{ \partial\hatN}{\partial \mt} \hatpb \barhatN 
      +\hatN  \hatpb \frac{ \partial \barhatN}{\partial \mt}  \right ) \right ]
  \Bigg  ].
\end{split} \label{eq:massShift}
\end{equation}

\section{The final result for the cross section }
\label{sect:final}

We now collect all the relevant formulae. We begin with the NLO cross section
expressed through the pole mass and write it in terms   of the short-distance mass
\begin{equation}
   \sigma  = \sigma_{\rm LO}(\mt) + \sigma_{R} +  \sigma_{V} +  \sigma_{\rm ren}
      =  \sigma_{\rm LO}({\tilde m}_t)  + \delta \sigma_{\rm NLO},
\end{equation}
where
\begin{equation}
\delta   \sigma_{\rm NLO} = \sigma_{R} +  \sigma_{V} + \sigma_{\rm ren} +
\delta  \sigma^{\rm expl}_{\rm mass} + \delta  \sigma^{\rm impl}_{\rm mass}.
\end{equation}

The individual contributions read 
\begin{equation}\label{eq:total}
  \begin{split}
    {\cal T}_\lambda  \left [  \delta  \sigma^{\rm expl}_{\rm mass} + \delta  \sigma^{\rm impl}_{\rm mass} \right ]
    & = \frac{\Cf \alpha_s}{2 \pi} \frac{\pi \lambda}{\mt}
    \; \int {\rm d Lips}_{\rm LO}  
    \times 
    \\
    & \;\;\;\;\;\;\Bigg [
    \frac{\mt^2}{\pd \ptop}
    \left [ 1 +  \pd^\mu \left ( \frac{\partial }{\partial \pd^\mu}
        - \frac{\partial }{\partial \ptop^\mu}
      \right ) \right ] F_{\rm LO}  -   \mt {\rm Tr} \left [ \hatone  \hatN \hatpb \barhatN \right ]
    \\
    & \;\;\;\;\;- \mt {\rm Tr} \left [ ( \hatptop + \mt) \left (
        \frac{ \partial\hatN}{\partial \mt} \hatpb \barhatN 
        +\hatN  \hatpb \frac{ \partial \barhatN}{\partial \mt}  \right ) \right ]
    \Bigg  ],
    \\
     {\cal T}_\lambda  \left[ \sigma_{\rm R} \right ] & =
    \frac{\alpha_s \Cf}{2 \pi }\frac{\pi \lambda}{\mt}
    \int  {\rm d Lips}_{\rm LO} 
    \Bigg [ 
    \left ( \frac{3}{2}  - \frac{ \mt^2}{  \pd  \ptop} - \frac{ \mt^2}{ \ptop \pb}  \right ) 
    \\
    & \;\;\;\;  - \frac{ \mt^2}{ \pd \ptop } { p}_d^\mu  \left ( \frac{\partial }{\partial { p}_d^\mu } - \frac{\partial }{\partial { p}_t^\mu}
    \right )  -  \frac{ \mt^2}{ { p}_t { p}_b} \pb^\mu  D_{p,\mu}  
    \Bigg  ] F_{\rm LO},
    \\
    {\cal T}_\lambda \left [  \sigma_{V}  \right ]
    & = 
    -\frac{\alpha_s \Cf}{2 \pi }\frac{\pi \lambda}{\mt}
    \int  {\rm d Lips}_{\rm LO} 
    \Bigg  [   {\rm Tr} \left [ \hatptop\hatN \hatpb \barhatN \right ]
    \\
    &
    \;\;\;\;\;+ \left (\frac{(2 \ptop \pb - \mt^2)}{\ptop \pb}
      - \frac{\mt^2}{\ptop \pb} \; \pb^\mu D_{p,\mu} \right ) F_{\rm LO}
    \Bigg  ],
    \\
    {\cal T}_\lambda \left [  \sigma_{\rm ren}  \right ]
    & = 
    \frac{\alpha_s \Cf}{2 \pi }\frac{\pi \lambda}{\mt}
    \int  {\rm d Lips}_{\rm LO} 
    \Bigg  [  \frac{3}{2} F_{\rm LO}
    \\
    &
    \;\;\;\;+
    \mt {\rm Tr} \left [ ( \hatptop + \mt) \frac{ \partial\hatN}{\partial \mt} \hatpb \barhatN \right ]
    +
    \mt {\rm Tr} \left [ ( \hatptop + \mt)\hatN  \hatpb \frac{ \partial \barhatN}{\partial \mt}  \right ]
    \Bigg  ].
 \end{split}
\end{equation}
Using the above results for the individual contributions, we obtain 
\begin{equation}
\begin{split} 
   {\cal T}_\lambda \left[ \delta  \sigma_{\rm NLO} \right ]
   & =
    \frac{\alpha_s \Cf}{2 \pi }\frac{\pi \lambda}{\mt}
        \int  {\rm d Lips}_{\rm LO} \;
        \left  (  F_{\rm LO} -  {\rm Tr} \left[  \hatptop \hatN \hatpb \barhatN \right ]
        - \mt {\rm Tr} \left[  \hatone\hatN \hatpb \barhatN \right ] \right ) =0.
\end{split}
\end{equation}
This result implies that ${\cal O}(\Lambda_{\rm QCD})$
corrections to processes  where single top quarks are produced by
virtue of weak flavor-changing interactions   vanish provided that the
cross section is expressed in terms of the short-distance top quark
mass.    In Appendix~\ref{app:SL}  we explain  how our method can be used to re-derive the known result  that there are no ${\cal O}(\Lambda_{\rm QCD})$ corrections
to semileptonic decays of a heavy quark~\cite{Bigi:1994em,Beneke:1994bc}.

\section{Alternative treatment of the self-energy corrections}\label{sect:altself}
The previous computation was first carried out in the pole-mass scheme, and then a scheme change
was performed to get the result in an arbitrary short distance scheme. Alternatively, it is possible
to perform the calculation directly in a short distance scheme.
In order to  do that,  we consider
 the squared amplitude directly  and recall that the external top quark  line is represented
by
\begin{equation}
  2\pi(\hatptop+\mt) \delta(\ptop^2-\mt^2) =   {\rm Disc}\left[ \frac{1}{\hatptop-\mt}\right] \equiv
 \left[ \frac{i}{\hatptop-\mt+i\epsilon}-\frac{i}{\hatptop-\mt-i\epsilon}   \right]\,,
\end{equation}
     One then deals with this external line in the same way as one deals with internal lines in Feynman diagrams, namely 
      one  inserts the self-energy correction and the mass counter-term into the argument of the ${\rm Disc}$ function,  but the wave function renormalisation does not need to be included. 
If the mass is renormalised in any short-distance scheme, we do not  need to include the mass
counter-term either, since it does not contain terms linear in $\lambda$. For the same reason,
mass counter-terms in the internal top quark lines are not needed. Thus, we can simply compute the self-energy
insertion without including any counter-term. The self-energy correction is given by
\begin{equation}
  \frac{i}{\hatptop -\mt}i\Sigma \frac{i}{\hatptop-\mt},
\end{equation}
where
\begin{equation}
  i\Sigma = \Cf\gs^2\int \frac{\mathd^4 k}{(2\pi)^4}
  \frac{-i}{k^2-\lambda^2+i\epsilon}(-i\gamma_\mu)\frac{i}{\hatptop-\hatk-\mt+i\epsilon}(-i\gamma^\mu).
\end{equation}
We need to evaluate $\Sigma$ up to terms that are suppressed  by more than one power of $\hatptop-\mt$, since higher
powers do not contribute to the discontinuity.
Making use of the virtual integrals given in Appendix~\ref{app:integrals}, a straightforward calculation
yields
\begin{equation}
 \begin{split}
 {\cal T}_\lambda  \left [  \Sigma  \right ]   &=\Cf \gs^2\left[\frac{1}{8\mt}(\ptop^2-\mt^2)+\frac{2\mt-\hatptop}{2}\right]\frac{1}{(2\pi)^2}
          \frac{\lambda \pi}{\sqrt{\ptop^2}} \\
        &= \frac{\as \Cf}{2\pi}\frac{\lambda\pi}{\mt}\left[-\frac{1}{4\mt}(\ptop^2-\mt^2)+2\mt-\hatptop\right]\,.
 \end{split}
 \end{equation}
 The full correction can be written as
\begin{equation}
 \begin{split}
  {\cal T}_{\lambda} \left [ {\rm Disc}\left[\left(\frac{i}{\hatptop-\mt}\right)^2 \Sigma\right] \right ]
  &=\frac{\as \Cf}{2\pi}\frac{\lambda\pi}{\mt}\,
    \Bigg[\frac{3}{2}(\hatptop+\mt)\,2\pi\,\delta(\ptop^2-\mt^2)
   \\ &
    -\mt 2\pi\,\delta(\ptop^2-\mt^2) +2\mt^2(\mt+\hatptop)\delta^\prime(\ptop^2-\mt^2)\Bigg], \label{eq:discinsertion}
 \end{split}
 \end{equation}
where $\delta^\prime(\ptop^2-\mt^2)$ is the derivative of the $\delta$-function with respect to $p_t^2$. 
In order to handle this derivative, we rewrite it as
\begin{equation}
  \begin{split} 
  \delta^\prime(\ptop^2-\mt^2)&=\frac{\pd^\mu}{2\pd  \ptop}\,\frac{\partial}{\partial \ptop^\mu}\delta(\ptop^2-\mt^2)
       =-\delta(\ptop^2-\mt^2) \frac{\partial}{\partial \ptop^\mu} \frac{\pd^\mu}{2\pd  \ptop}
  \\ &
       =\delta(\ptop^2-\mt^2)\left[\frac{4}{2\pd  \ptop}+\frac{\pd^\mu}{2(\pd  \ptop)^2}[-{p}_{t, \mu}+{p}_{b,\mu}]
       -\frac{\pd^\mu}{2\pd  \ptop}\frac{\partial}{\partial \ptop^\mu}\right]
  \\ &
       =\delta(\ptop^2-\mt^2)\left[\frac{3}{2\pd  \ptop}
         -\frac{\pd^\mu}{2\pd  \ptop}\frac{\partial}{\partial \ptop^\mu}\right]\,,
       \end{split}
\end{equation}
where we have  integrated  by parts, and we have assumed that in the phase space $\pd$ is taken as the dependent
momentum, i.e. $\pd=\pu+\pb-p_X-\ptop$. The remaining derivative with respect to $\ptop$ can be applied to
the amplitude or to the delta-function 
$\delta(\pd^2)$ in the phase space. In the second case we get
\begin{equation}
  -\frac{\pd^\mu}{2\pd  \ptop} \frac{\partial}{\partial \ptop^\mu} \; \delta(\pd^2)
  = \frac{1}{\pd  \ptop}\,\pd^2\delta^\prime(\pd^2)=-\frac{1}{2\pd  \ptop}\delta(\pd^2).
\end{equation}
Thus,   in eq.~(\ref{eq:discinsertion}) we can replace
\begin{equation}
   \delta^\prime(\ptop^2-\mt^2) \Rightarrow \delta(\ptop^2-\mt^2)\left[\frac{1}{2\pd  \ptop}
       -\frac{\pd^\mu}{2\pd  \ptop}\frac{\partial}{\partial \ptop^\mu}\right]\,, 
\end{equation}
with an understanding that the derivative acts only on the amplitude squared. Inserting eq.~(\ref{eq:discinsertion})
in the spinor trace, and including the phase space we get
\begin{equation}
  \delta\sigma_{\rm self}= \frac{\as \Cf}{2\pi} \frac{\lambda\pi}{\mt}\int \mathd {\rm Lips}_{\rm LO}
     \left[\frac{3}{2}F_{\rm LO}- \mt {\rm Tr}[ {\bf 1} \hatN\hatpb \barhatN]
       -\frac{\mt^2}{\pd {p}_t}\left(\pd^\mu\frac{\partial F_{\rm LO}}{\partial \ptop^\mu}-F_{\rm LO}\right)\right].
     \label{eq:selfencontr}
\end{equation}
In case $\pd$ is treated as an independent variable we must replace
\begin{equation}
  \frac{\partial}{\partial \ptop^\mu} \to \frac{\partial}{\partial \ptop^\mu} - \frac{\partial}{\partial \pd^\mu}\, , 
\end{equation}
and eq.~(\ref{eq:selfencontr}) becomes equivalent to the sum of the renormalisation contributions
of eq.~(\ref{eq:rencontr}) and the mass shift of eq.~(\ref{eq:massShift}).

\section{Kinematic distributions}
\label{sect:kinematics}

We will now study kinematic distributions in the single top production processes. We consider an
observable $X$ that depends on the momentum of the top quark
\begin{equation}
O_X = \int {\rm d} \sigma \; X(\qtop).
\label{eq7.1}
\end{equation}
To compute the ${\cal O}(\lambda)$ contribution to $O_X$, we follow the same route that was discussed in the previous
sections.  The difference with respect to the case of the inclusive cross section is the appearance of the  observable $X$
in the integrand in eq.~(\ref{eq7.1}). Remapping the momenta,  and  expanding  $X$ in  the gluon
momentum  $k$,  which appears in the argument of $X$ as the result of such remapping, we obtain
\begin{equation}
X(\qtop) = X(\ptop)
+  \frac{\partial X(\ptop)}{\partial \ptop^\mu } \left (
\frac{\ptop k}{\ptop \pd} \pd^\mu 
- k^\mu \right ).
\label{eq7.2}
\end{equation}

To compute the ${\cal O}(\lambda)$ contributions to $O_X$ it is convenient to combine the three terms in eq.~(\ref{eq7.2}) as follows
\be
   {\cal T}_\lambda [ O_X ] = {\cal T}_\lambda [ O^{(1)}_X ] + {\cal T}_\lambda [ O^{(2)}_X ],
\ee
where
\be
\begin{split} 
&    {\cal T}_\lambda [ O^{(1)}_X ]
    = {\cal T}_\lambda \left [ \int {\rm d} \sigma \; \left ( X(\ptop)
+  \frac{\partial X(\ptop)}{\partial \ptop^\mu } \; \frac{\ptop k}{\ptop \pd} \pd^\mu 
\right ) \right ],
\\
&    {\cal T}_\lambda [ O^{(2)}_X ]
    = -{\cal T}_\lambda \left [ \int {\rm d} \sigma \;  \frac{\partial X(\ptop)}{\partial \ptop^\mu } \; k^\mu  \right] .
\end{split} 
\ee

To compute ${\cal T}_\lambda [ O^{(1)}_X ]$, we note that   the observable $X(p_t)$ that appears there already depends
on the re-mapped momentum $\ptop$ and, for this reason, it does not affect the calculations
reported in the previous sections and the cancellation of ${\cal O}(\lambda)$ terms. The only
subtlety is that    the mass redefinition in  eq.~(\ref{eq:implicitmass}) produces an additional term
 because  in the current case  the derivative there must also act on  $X$.
 However, it is easy to see that this  new term  is
   \emph{exactly} compensated by the integral of the 
$k$-dependent term in the integrand of ${\cal T}_\lambda [ O^{(1)}_X ]$.  We conclude that
\be
{\cal T}_\lambda [ O_X^{(1)} ] = 0.
\ee

It remains to compute ${\cal T}_\lambda [ O_X^{(2)} ]$. Since the integrand is already proportional to $k$, we need the matrix element squared  and the
phase space in the leading soft approximation. We therefore find 
\begin{equation}
  {\cal T}_\lambda [ O_X^{(2)} ]
  = {\cal T}_\lambda\left[ C_F g_s^2 \int {\rm d} \sigma_{\rm LO} 
    \frac{\partial X(\ptop)}{\partial \ptop^\mu }
    \int \frac{ {\rm d}^4 k}{(2 \pi)^3} \,\delta_+(k^2-\lambda^2) \; J^\nu J_\nu\; k^\mu\right ],
\end{equation}
where  the eikonal current $J^\nu$ reads
\begin{equation}
  J^\nu\approx\frac{\ptop^\nu}{\ptop k}-\frac{\pb^\nu}{\pb k}.
\end{equation}
Using earlier  discussions  and the integrals presented in
Appendix~\ref{app:integrals}, it is straightforward to integrate this expression over $k$.
We obtain 
\begin{equation}
  {\cal T}_\lambda \left [ O_X  \right ] = {\cal T}_\lambda \left [ O_X^{(2)}  \right ]
  = \frac{\alpha_s \Cf}{2 \pi } \frac{\pi \lambda}{\mt}
  \; \int {\rm d} \sigma_{\rm LO}   \;
   l^\mu \frac{\partial X(\ptop)}{\partial \ptop^\mu }, \label{eq:diffdistrres}
\end{equation}
where
\begin{equation}
  l^\mu = \ptop^\mu -\frac{2 \mt^2}{\pb \ptop} \pb^\mu. 
\end{equation}

Using the alternative procedure for the inclusion of the self-energy corrections in Section~\ref{sect:altself} we
immediately reach the same conclusion, except that the cancellation of the second term of eq.~(\ref{eq7.2})
arises from the derivative term in eq.~(\ref{eq:selfencontr}) by replacing $F_{\rm LO}$ with $X\,F_{\rm LO}$,
 so that the derivative that hits $F_{\rm LO}$ there can  now also act on  $X(p_t)$.

The result of eq.~(\ref{eq:diffdistrres})
can be interpreted as a non-perturbative shift in the argument of the observable $X(\ptop)$. Indeed, we can write 
\begin{equation}
  \begin{split} 
    O_X &  = \int {\rm d} \sigma_{\rm LO} \left [ X(\ptop) 
      + \frac{\alpha_s \Cf}{2 \pi }
      \frac{\pi \lambda}{\mt}  \; l^\mu \frac{\partial X( \ptop)}{\partial \ptop^\mu}
    \right ]
    \\
    & = \int {\rm d} \sigma_{\rm LO} \;  X\left(\ptop+  \frac{\alpha_s \Cf}{2 \pi }
     \delta \ptop \right), 
  \end{split}
\end{equation}
where
\be
\delta  \ptop =\frac{\pi \lambda}{\mt}\;l.
\ee

As an example, suppose that $X$ is a function of the transverse
momentum distribution of the top quark, such as, for example, a cut on
the transverse momentum, or a product of theta functions singling out
a particular histogram bin. In this case
\begin{equation}
  {\ptop}_\perp  =\sqrt{| \ptop^\mu g_{\perp,\mu \nu} \ptop^\nu | },
\end{equation}
where
\begin{equation}
  g_{\perp}^{\mu \nu} = g^{\mu \nu} - \frac{\pb^\mu \pu^\nu + \pu^\mu  \pb^\nu }{\pu \pb}.
\end{equation}
Since $\pb^\mu g_{\perp,\mu \nu} =0$, we find 
\begin{equation}
  \frac{ \delta_{\rm NP} \left [\; {\ptop}_{\perp}  \right ] }{{\ptop}_{\perp}} = \frac{\alpha_s \Cf}{2 \pi }
  \frac{\pi \lambda}{\mt} .
\end{equation}

It is interesting to point out that the relative non-perturbative shift in ${\ptop}_{\perp}$  and the relative non-perturbative shift in the top quark mass coincide
\begin{equation}
  \frac{ \delta_{\rm NP} \left [ \; {\ptop}_{\perp}  \right  ] }{{\ptop}_{\perp}} = \frac{ \delta_{\rm  NP} [ m_t ]   }{m_t}.
\end{equation}
Since the non-perturbative uncertainty in the top quark mass is estimated as $100- 200~{\rm MeV}$
\cite{Beneke:2016cbu,Schwienhorst:2022yqu,Hoang:2017btd}, we conclude that the non-perturbative shift in the top quark
transverse momentum  reads 
\be
\delta_{\rm NP} \left [\; {\ptop}_{\perp}  \right ] \approx  (0.1-0.2)~\frac{{\ptop}_{\perp}}{m_t}~{\rm GeV}.  
\ee
The transverse momentum distribution of the $t$-channel  single top production is peaked around $50~{\rm GeV}$; for such momenta, the non-perturbative shift is very small, 
${\cal O}(30-60)$~MeV.

Another observable to consider is the top quark rapidity distribution. In the partonic center of mass frame, it reads 
\begin{equation}
  y_t = \frac{1}{2} \ln \frac{\pb \ptop }{\pu \ptop}.
\end{equation}
An easy computation gives
\begin{equation}
  \begin{split} 
    & \delta_{\rm NP} \left [ y_t \right ]
    = \frac{\alpha_s \Cf}{2 \pi }
    \frac{\pi \lambda}{\mt} \;  l^\mu \; \frac{1}{2} \left ( \frac{\pb^\mu}{\pb \ptop} - \frac{\pu^\mu}{\pu \ptop} \right )
    \\
    &    = \frac{\alpha_s \Cf}{2 \pi }
    \frac{\pi \lambda}{\mt} \frac{(\pu \pb) \mt^2}{(\pu \ptop) (\pb \ptop)} = 
    \frac{\alpha_s \Cf}{2 \pi }
    \frac{\pi \lambda}{\mt} \frac{8  \mt^2 s\, {\rm ch}^2(y_t)}{(s+\mt^2)^2}.
  \end{split} 
\end{equation}

\section{Conclusions}
\label{sect:concl}
   
In this paper we   discussed the non-perturbative  ${\cal O}(\Lambda_{\rm QCD})$ corrections
to
electroweak production of a single top quark in hadronic  collisions in
the context of renormalon calculus.  Processes of the type
$pp \to q + t + X$, where $X$ is an arbitrary collection of
colour-neutral particles, can be studied in the framework of renormalon
calculus because such processes do not contain gluons in leading order
diagrams.

We have shown how to use Low-Burnett-Kroll theorem, which allows one
to express sub-leading contributions in the soft expansion in a
process-independent way, to analyse ${\cal O}(\Lambda_{\rm QCD})$
corrections to \emph{arbitrary processes of a single top production
type}.  Our findings are remarkably simple. Indeed, we observe  that total cross
sections for such processes have no linear power corrections provided
that a short-distance mass scheme is  used to compute them. 
Therefore, if a total cross
section is employed  to determine the   top
quark mass,\footnote{For top quark pair production, this was   recently done in several experimental analyses and,
  at least in principle,  this can also be done for the single top production.}
it is more natural to use  a short-distance mass scheme since, by doing so, we avoid the presence of
linear renormalons.  Since renormalons are associated with the
factorial growth of the coefficients in perturbative series, the absence of linear renormalons should  lead to 
a better convergence of the perturbative expansion  in  a short-distance mass scheme. 
Although these conclusions    appear to be quite natural given what is known about semileptonic decays of heavy quarks,\footnote{
  Admittedly, this analogy cannot be complete since collider processes are not amenable to 
the  operator product expansion.}  our
calculation provides a strong indication  of the absence of ${\cal O}(\Lambda_{\rm QCD})$ corrections to  one of the main   top quark  production processes at a hadron
collider. Although  these results are obtained in the context of the renormalon calculus, we hope that they remain  valid also in full QCD.

We have also discussed how to generalise these results to compute
linear power corrections to kinematic distributions that involve the
top quark momentum. In this case, using  a short-distance mass scheme
and  making use of the pattern of
cancellations of various ${\cal O}(\lambda)$ contributions which
becomes apparent from the discussion of the total cross section, 
very simple formulae for ${\cal O}(\Lambda_{\rm QCD})$
non-perturbative shifts in the transverse momentum and rapidity
distributions of the top quark can be derived.

An important shortcoming of the approach to non-perturbative effects
in single top production developed in this paper is that it applies to
\emph{stable} top quarks. Since all ${\cal O}(\Lambda_{\rm QCD})$
corrections computed in this paper come from kinematic regions where
top quarks are nearly on shell, the instability of the top quark
should have a major effect on these results, suppressing linear power
corrections in realistic kinematic distributions.  In a related  context, an interplay between
the instability of the top quark and ${\cal O}(\Lambda_{\rm QCD})$ corrections were studied numerically
in ref.~\cite{FerrarioRavasio:2018ubr}.  In the future, it
would be interesting to investigate this interplay in more detail and establish
the degree of suppression of linear power corrections that otherwise  appear in
various kinematic distributions.

\section*{Acknowledgments}
We thank Adrian Signer for useful communications.
The research of K.M. was supported by the German Research Foundation (DFG, Deutsche Forschungsgemeinschaft) under grant 396021762-TRR 257.
P.~N. acknowledges the support of the Humboldt foundation.

\appendix

\section{Loop and real-emission integrals required for computing linear power corrections}\label{app:integrals}
   
 In this appendix  we present the results for the various integrals  that arise in the course of the
calculations reported in this paper.
To write the results for these  integrals  in a compact way, we introduce a variable 
\begin{equation}
\delta = \frac{1}{(2\pi)^2} \frac{\lambda \pi}{\mt}.
\end{equation}
\subsection{Real emission integrals}
The computation of the real emission integrals can be performed in the top quark rest frame, with an
arbitrary upper cutoff on the energy of the emitted gluon.  The result
does not depend upon the chosen frame, since the  only frame dependence
can  arise from the upper cutoff, and the soft region is not affected by
it. Thus one replaces
\begin{equation}
  \int \frac{{\rm d}^4 k}{(2\pi)^3} \delta_+(k^2 - \lambda^2)
  \Rightarrow \int_\lambda^{w_{\rm max}}
  \frac{\beta \omega \mathd \omega}{2(2\pi)^4} \int \mathd\varphi\,\int \mathd\cos\theta,
\end{equation}
where $\omega$ is the top quark energy, the polar axis is chosen along the direction of a $b$ quark 
and $\beta=\sqrt{1-{\lambda^2}/{\omega^2}}$, all in the top quark  rest frame.
All integrals are elementary; in the worst case one encounters integrals of the form
\begin{equation}
  \int_\lambda^{w_{\rm max}} \frac{\mathd \omega}{\omega^k}\,\log\frac{1+\beta}{1-\beta}
\end{equation}
that are easily done by parts, since
\begin{equation}
  \frac{\mathd}{\mathd\omega} \log\frac{1+\beta}{1-\beta}=\frac{2}{\sqrt{\omega^2-\lambda^2}}.
\end{equation}

The integrals required for computing the real emission contribution to single top production read\footnote{We only
  display ${\cal O}(\lambda)$ contributions to these integrals.}
\begin{align}
  & I_1 = {\cal T}_\lambda \left [  \int
    \frac{{\rm d}^4 k}{(2\pi)^3} \delta_+(k^2 - \lambda^2) \;  \frac{\lambda^2}{(2 \ptop k)^3}
    \right ]
    = \frac{1}{32 \mt^2} \; \delta,
  \\
  & I_2 = {\cal T}_\lambda \left [  \int \frac{{\rm d}^4 k}{(2\pi)^3} \delta_+(k^2 - \lambda^2) \;
    \frac{k^\mu  }{(2 \ptop k)^2}
    \right ]
    = -\frac{p^\mu_t}{8 \mt^2} \delta,
  \\
  & I_3 = {\cal T}_\lambda \left [  \int \frac{{\rm d}^4 k}{(2\pi)^3} \delta_+(k^2 - \lambda^2) \;
    \frac{1 }{(2 \ptop k)^2} 
    \right ]
    = 0,
  \\
  & I_4 = {\cal T}_\lambda \left [  \int \frac{{\rm d}^4 k}{(2\pi)^3} \delta_+(k^2 - \lambda^2) \;
    \frac{\lambda^2 }{(2 \ptop k)^2 (-2 \pb k)} 
    \right ]
    = -\frac{1}{16 (  \ptop \pb) } \delta,
  \\
  & I_5 = {\cal T}_\lambda \left [  \int \frac{{\rm d}^4 k}{(2\pi)^3} \delta_+(k^2 - \lambda^2) \;
    \frac{\lambda^2 }{(2 \ptop k) (-2 \pb k)^2} 
    \right ]
    = -\frac{\mt^2 }{16 ( \ptop \pb)^2} \delta,
  \\
  & I_6 = {\cal T}_\lambda \left [  \int \frac{{\rm d}^4 k}{(2\pi)^3} \delta_+(k^2 - \lambda^2) \;
    \frac{ k^\mu }{(2 \ptop k) (-2 \pb k)} 
    \right ]
    = \frac{1}{8 ( \ptop \pb)} \left ( \ptop^\mu - \frac{\mt^2 }{ \ptop \pb} \; \pb^\mu \right ) \delta.
\end{align}

\subsection{Loop integrals}
The required loop integrals read 
\begin{align}
  &  {\cal T}_\lambda \left [ 
    -i \int \frac{{\rm d}^4 k}{(2\pi)^4 (k^2 - \lambda^2)}
    J_t^\mu  J_{b,\mu} \right ] = \frac{1}{(4\pi)^2} \frac{\mt^2 - 2 \ptop \pb}{\ptop \pb } \frac{\pi \lambda}{\mt},
  \\
  & {\cal T}_\lambda \left [ 
    -i \int \frac{{\rm d}^4 k}{(2\pi)^4 (k^2 - \lambda^2)}
    J_t^\alpha J_{b,\alpha} k^\mu  \right ] = -\frac{2}{(4 \pi)^2} \frac{\pi \lambda}{\mt}
    \left ( \ptop^\mu - \frac{\mt^2 }{\ptop \pb} \pb^\mu \right ),
      \\
  & {\cal T}_\lambda \left [ 
    -i \int \frac{{\rm d}^4 k}{(2\pi)^4 (k^2 - \lambda^2)} J_t^\mu \right ] =
    -\frac{2}{(4 \pi)^2} \frac{\pi \lambda}{\mt} \; \ptop^\mu,
  \\
  & {\cal T}_\lambda \left [ 
    -i \int \frac{{\rm d}^4 k}{(2\pi)^4 (k^2 - \lambda^2)} J_b^\mu \right ] =0.
\end{align}

To compute them, we integrate over $k_0$ and map them onto real emission integrals.  More precisely, we 
first perform the replacement $k\rightarrow -k$ and then  perform  the $k^0$ integration in the $\ptop$ rest frame. The poles of the
$k^2-\lambda^2$, $d_b$ and $d_t$ denominators are given by
\begin{align}
  \omega &=\pm\sqrt{\vec{k}^2+\lambda^2}\mp i\epsilon, \label{eq:poles1} \\
  \omega &=p_b^0\pm\sqrt{(p_b^0)^2+(2\vec{k}  \vec{p}_b+\vec{k}^2)}\mp i\epsilon,
  \label{eq:poles2}\\
  \omega &=m\pm\sqrt{m^2+\vec{k}^2}\mp i\epsilon, \label{eq:poles3}
\end{align}
where $\omega=k^0$. We see that if we close the contour in the lower complex plane
we pick the residues of the poles with the upper signs in eqs.~(\ref{eq:poles1}-\ref{eq:poles3}),
i.e. the poles with negative imaginary part, but only the pole in eq.~(\ref{eq:poles1}) leads to a small value of $\omega$,
and thus leads to a term sensitive to $\lambda$.
Thus we can replace
\begin{equation}
  i  \int \frac{{\rm d}^4 k}{(2\pi)^4 (k^2 - \lambda^2)}
  \Rightarrow \int \frac{{\rm d}^4 k}{(2\pi)^3}\theta(k^0) \delta(k^2 - \lambda^2)\,,
\end{equation}
and then use the already known results for the real emission integrals.

\section{Semileptonic decays of a heavy quark}\label{app:SL}

In this section we consider the semileptonic decay of a top quark into a
massless bottom quark and an arbitrary collection of colour-neutral
particles, $t \to b + X$.  We will re-derive a well-known result \cite{Bigi:1994em,Beneke:1994bc} that
there are no ${\cal O}(\lambda)$ contributions to the total decay
width $\Gamma(t \to b +X)$ provided that the width is expressed in
terms of a short-distance mass of the top quark.

We note that all major steps of the calculation that we discussed in
the context of the single top production remain valid also for the
semileptonic decay. In particular, the calculation of the contribution of
the virtual corrections is identical.\footnote{Obviously, we need to
  account for the fact that in the decay process the top quark appears
  in the initial and the bottom quark in the final state.}  The
renormalisation procedure also remains the same.  As a result, we
find
\be
\begin{split}
  &  {\cal T}_\lambda \left [ m_t \; \Gamma_{V + \rm ren} \right ]
  = -\frac{\alpha_s C_F}{2 \pi} \frac{\pi \lambda}{m_t}
  \int {\rm dLips}(p_t|p_b,...)
  \Bigg  [   {\rm Tr} \left [ \hatpb\hatN \hatptop \barhatN \right ]
  \\
  &  -
  m_t {\rm Tr} \left [ \hatpb \frac{ \partial\hatN}{\partial m_t} ( \hatptop + m_t)   \barhatN \right ]
  -
  m_t {\rm Tr} \left [ \hatpb\hatN   ( \hatptop + m_t) \frac{ \partial \barhatN}{\partial m_t}  \right ]
  \\
  &  + \left (  -\frac{3}{2}  + \frac{(2 p_t p_b - m_t^2)}{p_t p_b}
    - \frac{m_t^2}{p_t p_b} \; p_b^\mu D_{p,\mu} \right ) F^{(d)}_{\rm LO}
  \Bigg  ],
\end{split}
\ee where
$F^{(d)}_{\rm LO} = {\rm Tr} \left [ \hatpb\hatN ( \hatptop + \mt)
  \barhatN \right ]$
is the leading order invariant amplitude squared.  We
also note that the above result is written for the product of the top
quark mass $m_t$ and the decay width, that is proportional to the
squared amplitude up to a numeric factor that is irrelevant for the
present purposes. We will see that it is $\Gamma$, rather than the
invariant amplitude, that is free of linear renormalons if expressed
in terms of a short-distance mass.

The calculation of the real-emission contributions proceeds similarly to the case of the single top production.
In particular, an application of Low-Burnett-Kroll theorem leads again to eq.~(\ref{eq2.22}) where for the decay
\be
\begin{split}
  & J^\mu = J_t^\mu + J_b^\mu,
  \\
  & J_t^\mu = \frac{2 p_t^\mu - k^\mu}{d_t},\;\;\;\; J_b^\mu = \frac{2 p_b^\mu + k^\mu}{d_b},
\end{split}
\ee
with $d_t = (p_t - k)^2 - m_t^2$ and $d_b = (p_2 + k)^2$.

In order to factorise the integration over the gluon
momentum from the rest of the phase space,
a momentum mapping is needed. This mapping differs from the one employed in
the discussion of the single top production.
We map the momentum of one of the colour-neutral, massless final-state particles (with momentum $p_3$)
and the $b$-quark as follows
\be
\begin{split}
  & p_b = \tilde p_b - k + \frac{ \tilde p_b k}{\tilde p_3 \tilde p_b} \tilde p_3,
  \;\;\;\; p_3 = \left (1 - \frac{\tilde p_b k}{\tilde p_3 \tilde p_b} \right ) \tilde p_3.
\end{split}
\ee
Upon this transformation, the phase space changes as follows
\be
   {\rm dLips}(p_t|p_b,p_3,k,...) = {\rm dLips}(p_t|\tilde p_b, \tilde p_3,..)
   \frac{{\rm d}^4 k }{(2\pi)^4} \delta(k^2 - \lambda^2)  \left ( 1 + \frac{k \tilde p_3}{\tilde p_b \tilde p_3}
   - \frac{k \tilde p_b }{\tilde p_b \tilde p_3}
   \right ).
\ee
Integrating over the gluon momentum $k$ using the integrals in Appendix~\ref{app:integrals},
we obtain the  real emission contribution
\be
\begin{split}
  & {\cal T}_\lambda  \left[ m_t \; \Gamma_{\rm R} \right ]=
  \frac{\alpha_s C_F}{2 \pi }\frac{\pi \lambda}{m_t}
  \int  {\rm d Lips}(p_t,\tilde p_b, \tilde p_3, p_X)
  \Bigg [
  \left ( \frac{1}{2}  - \frac{\tilde p_3 p_t }{\tilde p_3 \tilde p_b}
    +  \frac{\tilde p_b p_t }{\tilde p_3 \tilde p_b} - \frac{m_t^2}{\tilde p_b p_t}
  \right )
  \\
  &
  -\frac{\tilde p_b p_t}{\tilde p_3 \tilde p_b} {\tilde p}_3^\mu \left ( \frac{\partial }{\partial {\tilde p}_b^\mu} - \frac{\partial }{\partial {\tilde p}_3^\mu} \right )
  -\frac{m_t^2}{\tilde p_b p_t} \tilde p_b^\mu D_\mu
  + p_t^\mu D_\mu
  \Bigg  ] F_{\rm LO},
\end{split}
\ee

The most important difference in comparison with the single top
production computation comes from the change in the cross section due
to the mass redefinition  since in the current  case the top quark is in the
initial state. Nevertheless, it is possible to change the quark mass redefining momenta. We write
\be
\begin{split}
  p_b^\mu = \tilde p_b^\mu - \kappa \tilde p_t + \kappa \frac{\tilde p_b \tilde p_t}{ \tilde p_b \tilde p_3} \tilde p_3^\mu,
  \;\;\;\; p_3^\mu = \tilde p_3^\mu \left ( 1 - \kappa \frac{\tilde p_b \tilde p_t}{ \tilde p_b \tilde p_3} \right ),\;\;\; p_t^\mu=\tilde p_t^\mu (1-\kappa).
\end{split}
\ee
The phase space becomes
\be
   {\rm dLips}(p_t|p_b,p_3,...) = {\rm dLips}(\tilde p_t |\tilde p_b, \tilde p_3,..)
   \left ( 1 + \kappa \frac{ \tilde p_t \tilde p_3}{\tilde p_b \tilde p_3}
   - \kappa \frac{ \tilde p_b \tilde p_t }{\tilde p_b \tilde p_3},
   \right ).
\ee
Choosing   $\kappa = C_F \alpha_s/(2\pi) \pi \lambda/m_t$,  we find that
$\tilde p_t^2$ corresponds to the short-distance mass $\tildemt$ defined in eq.~(\ref{eq5.8a}).

Similar to the case of single top production, we need to consider the changes in leading
order width due to explicit and implicit mass redefinitions.   We write
\be
 m_t \Gamma^{\rm LO}(m_t) -  \tilde m_t \Gamma^{\rm LO}({\tilde m}_t) = \delta  [ m_t \Gamma]_{\rm impl} + \delta [m_t \Gamma]_{\rm expl}.
\ee
The implicit change is caused by changing the top quark mass  in phase space; we account for this using momenta redefinitions described above.
We find
\be
\begin{split}
\delta \left [ m_t \Gamma \right ]_{\rm impl}
  &  = \frac{C_F \alpha_s}{2 \pi} \frac{\pi \lambda}{m_t} \; \int {\rm dLips}(\tilde p_t |\tilde p_b, \tilde p_3,..)
   \Bigg  [
    \frac{ \tilde p_t \tilde p_3}{\tilde p_b \tilde p_3}
   -  \frac{ \tilde p_b \tilde p_t }{\tilde p_b \tilde p_3}
   - \tilde p_t^\mu D_\mu
\\
&  + \frac{\tilde p_b \tilde p_t}{\tilde p_b \tilde p_3} \tilde p_3^\mu \left ( \frac{\partial}{\partial \tilde p_b^\mu}
- \frac{\partial }{\partial \tilde p_3^\mu} \right )
   \Bigg  ] F_{\rm LO}.
\end{split}
     \ee
     In addition, there is an explicit change in leading order width related to a replacement of the mass $m_t$ in the amplitude.
     We find
\be
\begin{split}
\delta \left [ m_t  \Gamma \right ]_{\rm expl}
   & = -\frac{C_F \alpha_s}{2 \pi} \frac{\pi \lambda}{m_t} \; m_t \int {\rm dLips}(\tilde p_t |\tilde p_b, \tilde p_3,..)
   \Bigg [
      {\rm Tr} \left [ \hatpb \hatN  \hatone     \barhatN \right ]
\\
&      + {\rm Tr}\left [ \hatpb \frac{\partial }{\partial m_t}\hatN ( \hatptop + m_t ) \barhatN
       + \hatpb \hatN ( \hatptop + m_t ) \frac{\partial }{\partial m_t} \barhatN \right ]
       \Bigg  ].
\end{split}
\ee

We define the correction to the width $\Gamma_{\rm NLO}$ through the following formula
\be
\Gamma_{\rm LO}(m_t)  + \Gamma_{V+\rm ren} + \Gamma_{R} = \Gamma_{\rm LO}(\tilde m_t) + \Gamma_{\rm NLO}.
\ee
Writing
\be
\begin{split}
   {\cal T}_\lambda \left [ \Gamma_{\rm NLO} \right ]
  &  = \frac{1}{m_t} \Bigg  [
     \delta  [ m_t \Gamma]_{\rm impl} + \delta [m_t \Gamma]_{\rm expl}
     + (\tilde m_t - m_t ) \Gamma_{\rm LO}
     \\
     & + {\cal T}_\lambda \left [ m_t  \Gamma_{V+\rm ren} \right ] + {\cal T}_\lambda \left [ m_t \Gamma_R \right ]
     \Bigg  ],
   \end{split} 
\ee
and using explicit expressions for the various contributions on the right hand side of the above equation, we obtain the well-known result
\be
{\cal T}_\lambda \left [ \Gamma_{\rm NLO} \right ] = 0.
\ee

\bibliographystyle{JHEP}
\bibliography{sintop}

\providecommand{\href}[2]{#2}\begingroup\raggedright\begin{thebibliography}{10}

\bibitem{Schwienhorst:2022yqu}
K.~Agashe et~al., \emph{{Report of the Topical Group on Top quark physics and
  heavy flavor production for Snowmass 2021}},
  \href{http://arxiv.org/abs/2209.11267}{{\tt 2209.11267}}.

\bibitem{Azzi:2019yne}
P.~Azzi et~al., \emph{{Report from Working Group 1}: {Standard Model Physics at
  the HL-LHC and HE-LHC}},
  \href{http://dx.doi.org/10.23731/CYRM-2019-007.1}{\emph{CERN Yellow Rep.
  Monogr.} {\bf 7} (2019) 1--220}, [\href{http://arxiv.org/abs/1902.04070}{{\tt
  1902.04070}}].

\bibitem{Bigi:1994em}
I.~I.~Y. Bigi, M.~A. Shifman, N.~G. Uraltsev and A.~I. Vainshtein, \emph{{The
  Pole mass of the heavy quark. Perturbation theory and beyond}},
  \href{http://dx.doi.org/10.1103/PhysRevD.50.2234}{\emph{Phys. Rev. D} {\bf
  50} (1994) 2234--2246}, [\href{http://arxiv.org/abs/hep-ph/9402360}{{\tt
  hep-ph/9402360}}].

\bibitem{Beneke:1994sw}
M.~Beneke and V.~M. Braun, \emph{{Heavy quark effective theory beyond
  perturbation theory: Renormalons, the pole mass and the residual mass term}},
  \href{http://dx.doi.org/10.1016/0550-3213(94)90314-X}{\emph{Nucl. Phys. B}
  {\bf 426} (1994) 301--343}, [\href{http://arxiv.org/abs/hep-ph/9402364}{{\tt
  hep-ph/9402364}}].

\bibitem{Langenfeld:2009wd}
U.~Langenfeld, S.~Moch and P.~Uwer, \emph{{Measuring the running top-quark
  mass}}, \href{http://dx.doi.org/10.1103/PhysRevD.80.054009}{\emph{Phys. Rev.
  D} {\bf 80} (2009) 054009}, [\href{http://arxiv.org/abs/0906.5273}{{\tt
  0906.5273}}].

\bibitem{Dowling:2013baa}
M.~Dowling and S.-O. Moch, \emph{{Differential distributions for top-quark
  hadro-production with a running mass}},
  \href{http://dx.doi.org/10.1140/epjc/s10052-014-3167-x}{\emph{Eur. Phys. J.
  C} {\bf 74} (2014) 3167}, [\href{http://arxiv.org/abs/1305.6422}{{\tt
  1305.6422}}].

\bibitem{Makela:2023wbk}
T.~M\"akel\"a, A.~Hoang, K.~Lipka and S.-O. Moch, \emph{{Investigation of the
  scale dependence in the MSR and $\overline{\textrm{MS}}$ top quark mass
  schemes for the $\mathrm{t}\overline{\mathrm{t}}$ invariant mass differential
  cross section using LHC data}},  \href{http://arxiv.org/abs/2301.03546}{{\tt
  2301.03546}}.

\bibitem{Caola:2021kzt}
F.~Caola, S.~Ferrario~Ravasio, G.~Limatola, K.~Melnikov and P.~Nason, \emph{{On
  linear power corrections in certain collider observables}},
  \href{http://dx.doi.org/10.1007/JHEP01(2022)093}{\emph{JHEP} {\bf 01} (2022)
  093}, [\href{http://arxiv.org/abs/2108.08897}{{\tt 2108.08897}}].

\bibitem{FerrarioRavasio:2018ubr}
S.~Ferrario~Ravasio, P.~Nason and C.~Oleari, \emph{{All-orders behaviour and
  renormalons in top-mass observables}},
  \href{http://dx.doi.org/10.1007/JHEP01(2019)203}{\emph{JHEP} {\bf 01} (2019)
  203}, [\href{http://arxiv.org/abs/1810.10931}{{\tt 1810.10931}}].

\bibitem{Beneke:1998ui}
M.~Beneke, \emph{{Renormalons}},
  \href{http://dx.doi.org/10.1016/S0370-1573(98)00130-6}{\emph{Phys. Rept.}
  {\bf 317} (1999) 1--142}, [\href{http://arxiv.org/abs/hep-ph/9807443}{{\tt
  hep-ph/9807443}}].

\bibitem{Low:1958sn}
F.~E. Low, \emph{{Bremsstrahlung of very low-energy quanta in elementary
  particle collisions}},
  \href{http://dx.doi.org/10.1103/PhysRev.110.974}{\emph{Phys. Rev.} {\bf 110}
  (1958) 974--977}.

\bibitem{Burnett:1967km}
T.~H. Burnett and N.~M. Kroll, \emph{{Extension of the low soft photon
  theorem}}, \href{http://dx.doi.org/10.1103/PhysRevLett.20.86}{\emph{Phys.
  Rev. Lett.} {\bf 20} (1968) 86}.

\bibitem{Engel:2021ccn}
T.~Engel, A.~Signer and Y.~Ulrich, \emph{{Universal structure of radiative QED
  amplitudes at one loop}},
  \href{http://dx.doi.org/10.1007/JHEP04(2022)097}{\emph{JHEP} {\bf 04} (2022)
  097}, [\href{http://arxiv.org/abs/2112.07570}{{\tt 2112.07570}}].

\bibitem{Akhoury:1996ks}
R.~Akhoury, L.~Stodolsky and V.~I. Zakharov, \emph{{Power corrections and KLN
  cancellations}},
  \href{http://dx.doi.org/10.1016/S0550-3213(97)00766-9}{\emph{Nucl. Phys. B}
  {\bf 516} (1998) 317--332}, [\href{http://arxiv.org/abs/hep-ph/9609368}{{\tt
  hep-ph/9609368}}].

\bibitem{Akhoury:1997pb}
R.~Akhoury, M.~G. Sotiropoulos and V.~I. Zakharov, \emph{{The KLN theorem and
  soft radiation in gauge theories: Abelian case}},
  \href{http://dx.doi.org/10.1103/PhysRevD.56.377}{\emph{Phys. Rev. D} {\bf 56}
  (1997) 377--387}, [\href{http://arxiv.org/abs/hep-ph/9702270}{{\tt
  hep-ph/9702270}}].

\bibitem{Beneke:1994bc}
M.~Beneke, V.~M. Braun and V.~I. Zakharov, \emph{{Bloch-Nordsieck cancellations
  beyond logarithms in heavy particle decays}},
  \href{http://dx.doi.org/10.1103/PhysRevLett.73.3058}{\emph{Phys. Rev. Lett.}
  {\bf 73} (1994) 3058--3061}, [\href{http://arxiv.org/abs/hep-ph/9405304}{{\tt
  hep-ph/9405304}}].

\bibitem{tHooft:1973mfk}
G.~'t~Hooft, \emph{{Dimensional regularization and the renormalization group}},
  \href{http://dx.doi.org/10.1016/0550-3213(73)90376-3}{\emph{Nucl. Phys. B}
  {\bf 61} (1973) 455--468}.

\bibitem{Czarnecki:1997sz}
A.~Czarnecki, K.~Melnikov and N.~Uraltsev, \emph{{Non-Abelian dipole radiation
  and the heavy quark expansion}},
  \href{http://dx.doi.org/10.1103/PhysRevLett.80.3189}{\emph{Phys. Rev. Lett.}
  {\bf 80} (1998) 3189--3192}, [\href{http://arxiv.org/abs/hep-ph/9708372}{{\tt
  hep-ph/9708372}}].

\bibitem{Beneke:1998rk}
M.~Beneke, \emph{{A Quark mass definition adequate for threshold problems}},
  \href{http://dx.doi.org/10.1016/S0370-2693(98)00741-2}{\emph{Phys. Lett. B}
  {\bf 434} (1998) 115--125}, [\href{http://arxiv.org/abs/hep-ph/9804241}{{\tt
  hep-ph/9804241}}].

\bibitem{Hoang:1999ye}
A.~H. Hoang, \emph{{1S and MS-bar bottom quark masses from Upsilon sum rules}},
  \href{http://dx.doi.org/10.1103/PhysRevD.61.034005}{\emph{Phys. Rev. D} {\bf
  61} (2000) 034005}, [\href{http://arxiv.org/abs/hep-ph/9905550}{{\tt
  hep-ph/9905550}}].

\bibitem{Hoang:2008yj}
A.~H. Hoang, A.~Jain, I.~Scimemi and I.~W. Stewart, \emph{{Infrared
  Renormalization Group Flow for Heavy Quark Masses}},
  \href{http://dx.doi.org/10.1103/PhysRevLett.101.151602}{\emph{Phys. Rev.
  Lett.} {\bf 101} (2008) 151602}, [\href{http://arxiv.org/abs/0803.4214}{{\tt
  0803.4214}}].

\bibitem{Beneke:2016cbu}
M.~Beneke, P.~Marquard, P.~Nason and M.~Steinhauser, \emph{{On the ultimate
  uncertainty of the top quark pole mass}},
  \href{http://dx.doi.org/10.1016/j.physletb.2017.10.054}{\emph{Phys. Lett. B}
  {\bf 775} (2017) 63--70}, [\href{http://arxiv.org/abs/1605.03609}{{\tt
  1605.03609}}].

\bibitem{Hoang:2017btd}
A.~H. Hoang, C.~Lepenik and M.~Preisser, \emph{{On the Light Massive Flavor
  Dependence of the Large Order Asymptotic Behavior and the Ambiguity of the
  Pole Mass}}, \href{http://dx.doi.org/10.1007/JHEP09(2017)099}{\emph{JHEP}
  {\bf 09} (2017) 099}, [\href{http://arxiv.org/abs/1706.08526}{{\tt
  1706.08526}}].

\end{thebibliography}\endgroup

\end{document}